\documentclass[twocolumn,showpacs,preprintnumbers,amsmath,amssymb]{revtex4-1}
\usepackage{textcomp}
\usepackage[pdftex]{epsfig}
\DeclareGraphicsExtensions{.jpg,.mps,.pdf,.png}

\usepackage{dcolumn}
\usepackage{bm}
\usepackage{comment}
\usepackage{color}

\newcommand{\jcap}{{\em Journal of Cosmology and Astro-Particle Physics\/} }
\newcommand{\apjl}{{\em Astrophys. J. Letter\/}}

\newcommand{\mnras}{{\em Mon. Not. R. Astr. Soc. \/}}

\newcommand{\physrep}{{\em Phys. Rep. \/}}

\newcommand{\fw}{w}
\newcommand{\ini}{{\rm in}}

\newcommand{\vd}{{\bf d}}

\newcommand{\vq}{{\bf q}}

\newcommand{\vv}{{\bf v}}

\newcommand{\vk}{{\bf k}}
\newcommand{\dd}{{\rm d}}

\newcommand{\mg}{\big<}
\newcommand{\md}{\big>}

\newcommand{\mP}{{\cal P}}

\newcommand{\cdm}{{\rm c}}
\newcommand{\ba}{{\rm b}}

\newcommand{\hxi}{{\hat{\xi}}}
\newcommand{\hP}{{\hat{P}}}
\newcommand{\hdelta}{{\hat{\delta}}}

\newcommand{\Dirac}{\delta_{\rm D}}

\newcommand{\be}{\[}
\newcommand{\ee}{\]}

\newcommand{\beqa}{\begin{eqnarray}}
\newcommand{\eeqa}{\end{eqnarray}}

\renewcommand{\[}{\begin{equation}}
\renewcommand{\]}{\end{equation}}

\def\la{\mathrel{\mathpalette\fun <}}
\def\ga{\mathrel{\mathpalette\fun >}}
\def\fun#1#2{\lower3.6pt\vbox{\baselineskip0pt\lineskip.9pt
        \ialign{$\mathsurround=0pt#1\hfill##\hfil$\crcr#2\crcr\sim\crcr}}}

\newcommand{\rhob}{\overline{\rho}}

\newcommand{\bdm}{\begin{displaymath}}
\newcommand{\edm}{\end{displaymath}}
\newcommand{\bea}{\begin{eqnarray}}
\newcommand{\eea}{\end{eqnarray}}
\newcommand{\bt}{\begin{tabular}}
\newcommand{\et}{\end{tabular}}

\newcommand{\eP}{P_{\rm eff}}
\newcommand{\DP}{\Delta \hat P}

\newcommand{\HH}{{\cal H}}

\newcommand{\rtheta}{{\Theta}}

\newcommand{\egamma}{\gamma}

\newcommand{\rcomp}{{\cal R}}
\newcommand{\ircomp}{{\cal S}}

\newcommand{\Mpc}{{\rm Mpc}}
\newcommand{\Kpc}{{\rm kpc}}
\newcommand{\cis}{{\triangleleft}}
\newcommand{\iis}{{\triangledown}}
\newcommand{\plus}{{+}}
\newcommand{\minus}{{-}}

\newcommand{\ci}{{\cis}}
\newcommand{\vdi}{{\bf d^{\iis}}}
\newcommand{\hvdi}{{\hat {\bf d}^{\iis}}}
\newcommand{\ndi}{{d^{\iis}}}

\newcommand{\bb}{I}
\newcommand{\bc}{J}

\newcommand\hTheta{\hat{\Theta}}
\newcommand\ialpha{\hat\alpha}
\newcommand{\deltam}{\delta_{\rm m}}
\newcommand{\ii}{{\rm i}}

\begin{document}

\title{Power spectra in the eikonal approximation with adiabatic and nonadiabatic modes}

\author{Francis Bernardeau, Nicolas Van de Rijt, Filippo Vernizzi}
\affiliation{Institut de Physique Th{\'e}orique,
         CEA/DSM/IPhT, Unit{\'e} de recherche associ{\'e}e au CNRS, CEA/Saclay
         91191 Gif-sur-Yvette c{\'e}dex}
\vspace{.2 cm}
\date{\today}
\vspace{.2 cm}
\begin{abstract}
We use the so-called eikonal approximation, recently introduced in the context of cosmological perturbation theory, to compute power spectra for multi-component fluids. We demonstrate that, at any given order in standard perturbation theory, multipoint power spectra do not depend on the large-scale adiabatic modes.
Moreover, we employ perturbation theories to decipher how nonadiabatic modes, such as a relative velocity between two different components, damp the small-scale matter power spectrum, a mechanism recently described in the literature. 
In particular, we do an explicit calculation at 1-loop order of this effect. 
 While the 1-loop result eventually breaks down, we show how the damping effect
can be fully captured by the help of the eikonal approximation. A relative velocity not only induces mode damping but also creates large-scale anisotropic modulations of the matter power spectrum  amplitude. We illustrate this  for  the Local Group environment.
\end{abstract}
 \vskip2pc

\maketitle

\section{Introduction}

Recently, there has been   much progress in Perturbation Theory (PT) calculations that have greatly improved our understanding of gravitational instability beyond the linear regime. In particular, it has been recognized that there exist many alternative 
resummmation schemes to the standard PT, such as those proposed in \cite{Crocce:2005xy,Crocce:2005xz,Crocce:2007dt,Matsubara:2007wj, Matsubara:2008wx,McDonald:2006hf,Izumi:2007su,Taruya:2007xy,Taruya:2009ir, Pietroni:2008jx,Matarrese:2007wc,Valageas:2003gm, Valageas:2006bi,2012arXiv1207.1465C,2012arXiv1208.1191T,2012JCAP...04..013T,2012arXiv1206.2926C}, that aim at improving upon its sometimes poor convergence (see, e.g., 
\cite{Crocce:2005xy,Carlson:2009it,Taruya:2009ir}).
A genuine breakthrough emerged with Refs.~\cite{Crocce:2005xy,Crocce:2005xz,2008PhRvD..78j3521B},  where it was shown that some classes of next-to-leading order corrections in PT could be explicitly resummed. These references also showed that the so-called propagators, connecting the final state of a perturbation variable to the initial conditions, are damped by the nonlinear coupling  with long-wavelength modes, assuming that the latter are in the linear regime and obey Gaussian statistics. Although these results were initially derived in a specific context, they proved to be very generic and have later been extended to more elaborated types of objects and to arbitrary initial conditions \cite{2008PhRvD..78j3521B,2010PhRvD..82h3507B}. In \cite{2012PhRvD..85f3509B}---although the original ideas can be already found in \cite{2010PhRvD..82h3507B,2008PhRvD..78h3503B,2007arXiv0711.3407V}---we have shown  that they can be entirely captured with the help of the so-called \textsl{eikonal approximation}. 

The eikonal approach allows us to take into account the effect of large-scale modes on the development of small-scale perturbations in a non-perturbative way. In essence, it is based on a separation of scales. This is not new: as it is often the case in cosmological studies, the idea is to compute the impact of large-scale modes on the small-scale dynamics. If long wavelengths are much longer than the short ones and are  \textsl{adiabatic}, their effect can be seen as a redefinition of the \textsl{background} for the small scales.
For instance, such a heuristic approach has been employed to compute the bias functions (see e.g.~\cite{2002PhR...372....1C} for a review).

Also in the eikonal approximation the separation of scales is used to redefine a new set of equations governing the small-scale physics when this is embedded in an environment shaped by large-scale cosmological modes. However, this is based on a well-defined scheme in PT; thus,  at variance to most cases in cosmology where the long modes are implicitly or explicitly assumed to be adiabatic, the eikonal approach can also incorporate
nonadiabatic large-scale modes appearing in a multi-fluid context. 
Indeed, in \cite{2012PhRvD..85f3509B} we have applied this approach to a system of multiple (pressureless) fluids and demonstrated that the development of gravitational instabilities could greatly be affected by the presence of \textsl{isodensity} modes. In particular, this can be the case for a cosmology with baryons and Cold Dark Matter (CDM) particles.

In \cite{2012PhRvD..85f3509B}, applications of the eikonal approximation  focused on the behavior of (multipoint) propagators, which can be seen as the building blocks for constructing PT predictions at next-to-leading orders. In this paper we rather concentrate our calculations on  power spectra amplitudes.
In particular, we first wish to clarify the role of adiabatic large-scale perturbations on small-scale power spectra. Indeed, we will show at any order in standard PT that power spectra and multipoint spectra \textsl{do not} depend on large-scale adiabatic modes.

Secondly, we intend to make contact with the effect recently described in \cite{2010PhRvD..82h3520T}, where it was shown that a relative velocity between  baryons and CDM  could affect the growth of structures  on small scales and damp their amplitude at high redshift.
One of the aims of this paper is to rederive this effect in the framework of standard PT and show that it is naturally captured by the theoretical setting provided by the eikonal approximation. 

The impact of the relative velocity between baryons and CDM on the formation of first stars and early structures 
has been further investigated analytically in \cite{2011MNRAS.418..906T,Bovy:2012af} and, using N-body simulations and semianalytical codes, in \cite{2011MNRAS.412L..40M,2011ApJ...730L...1S,2011ApJ...736..147G,2011arXiv1110.2111F,Bittner:2011rx,McQuinn:2012rt}.
In \cite{2010JCAP...11..007D} it is claimed that this effect imprints a characteristic shape on the power spectrum at baryonic acoustic oscillation scales, which can be used to prove physics on kpc scales, while the authors of \cite{2011JCAP...07..018Y} discuss how to accurately take into account  the impact of this effect  in baryonic acoustic oscillation measurements. The effect on the cosmic microwave background $B$-mode polarization has also been studied in \cite{Ferraro:2011nc,Grin:2011nk}.

The plan of the paper is the following. As some of the effects that we want to study are engrained into multiple fluid physics, we first review  the dynamical equations describing a set of noninteracting pressureless fluids in Sec.~\ref{sec:MultiPT} and introduce the concepts of propagators and of adiabatic and isodensity modes. This provides us with the necessary ingredients to compute the 1-loop correction to the power spectrum. In particular, we will focus on the small-scale behavior of the power spectrum, showing that at leading order the large-scale adiabatic modes do not affect the 1-loop correction, while the large-scale isodensity modes do so, 
leading to damping with uncontrolled divergences. 
In Sec.~\ref{sec:SectEikonal}  we review the eikonal approximation. We then present one of the key results of this paper, i.e.~the fact that large-scale adiabatic modes do not
contribute to power spectra on small scales, at any order in standard PT calculation. Finally,  in the subsequent section (Sec.~\ref{sec:SectionIV}) we focus our attention on the effect of isodensity modes in the context of the eikonal approximation making contact with the behavior uncovered in Sec.~\ref{sec:MultiPT}. 
More specifically, the divergences found at 1-loop order are here regularized. 
We then present the expected shape of the small-scale matter power spectrum due to isodensity modes and we illustrate these results for the Milky Way environment.
Astrophysical consequences are discussed in the conclusion.

\section{Perturbation Theory in a multiple-component fluid}
\label{sec:MultiPT}

In this section we review the formalism for PT calculations in the presence of multiple pressureless fluids and we extend  some of the standard PT calculations described in \cite{2002PhR...367....1B} to the case of two  fluids. 
The equations modeling multiple pressureless fluids have already been presented in previous papers,  e.g.~\cite{2010PhRvD..81b3524S,2012PhRvD..85f3509B}. Later on, in Sec.~\ref{sec:SectionIV}, we will apply them to baryons and CDM and only then, in order to make more realistic predictions, we will introduce a pressure gradient term for the baryonic component.

\subsection{Equations of motion and  linear solutions}

We assume that the Universe is filled with pressureless fluids with only gravitational interactions.
For $N$ fluids, we  denote each fluid by a subscript $\bb$ ($\bb=1,\ldots, N$).
Then, for each fluid the continuity equation reads
\[
\frac{\partial}{\partial t} \delta_{\bb}+\frac1a \left((1+\delta_{\bb})u_{\bb}^{i}\right)_{,i}=0\;,
\]
while the Euler equation reads
\[
\frac{\partial}{\partial t}u_{\bb}^{i}+H u_{\bb}^{i}+\frac{1}{a}u_{\bb}^{j}u_{\bb,j}^{i}=-\frac{1}{a}\phi_{,i}\;.
\]
The Poisson equation,
\[
\Delta \phi=4\pi\,G\,a^{2}\rhob\,\deltam,
\]
where $\deltam$ is the density contrast of the total fluid energy density, i.e.
\[
\rho \equiv \sum_\bb \rho_{\bb} \equiv  (1+\deltam) \rhob\;,
\]
allows us to close the system. In Fourier space, the equations of motion become
\begin{align}
\frac{1}{H}\frac{\partial}{\partial t}\delta_{\bb}(\vk)+\theta_{\bb}(\vk)= - \alpha(\vk_{1},\vk_{2})\theta_{\bb}(\vk_{1})\delta_{\bb}(\vk_{2}) \;,
\label{Cont2}\\
\frac{1}{H}\frac{\partial}{\partial t}\theta_{\bb}(\vk)+\left(2+\frac{1}{H^2}\frac{\dd H}{\dd t}\right)\theta_{\bb}(\vk)+\frac{3}{2}\Omega_{\rm m}\deltam(\vk)=
\nonumber\\
\qquad\qquad- \beta(\vk_{1},\vk_{2})\theta_{\bb}(\vk_{1})\theta_{\bb}(\vk_{2}) \;,&\label{Eul2}
\end{align}
where  a convolution on the right-hand side of this equation is implicitly assumed. Here $\theta_{\bb}$ is the dimensionless divergence of the velocity field of the fluid $\bb$,  the coupling functions are given by
\begin{align}
\alpha(\vk_{1},\vk_{2}) &=\frac{(\vk_{1}+\vk_{2})\cdot \vk_{1}}{k_{1}^2} \;, \label{alpha} \\
\beta(\vk_{1},\vk_{2}) &=\frac{(\vk_{1}+\vk_{2})^2\ \vk_{1}\cdot \vk_{2}}{2k_{1}^2 k_{2}^2}\;, \label{beta}
\end{align}
and  $\Omega_{\rm m}$ is the reduced total density of the pressureless fluids. The mutual  coupling between the two fluids is due to the gravitational force in the Euler equation, which by the Poisson equation is proportional to $\deltam$.

The linear solutions of this system of equations have been presented in \cite{2001NYASA.927...13S,2010PhRvD..81b3524S,2012PhRvD..85f3509B}, where it is shown that it is convenient to introduce the multiplet $\Psi_a$ ($a=1,\ldots, 2N$),
\[
\Psi_{a}=\left(
\delta_{1},
\rtheta_{1},
\delta_{2},
\rtheta_{2},
\dots
\right)^T \;, \label{Psi_def_2}
\]
with $
\rtheta_{\bb} \equiv - {\theta_{\bb}}/{f_{+}(t)}$. The growth rate $f_{+}$ is defined as the logarithmic change of the growth factor with the expansion, $f_{+} \equiv \dd \ln D_+/ \dd \eta$, with $\dd \eta \equiv \dd \ln a$.
The equations of motion can be recapped in the form
\[
\frac{\partial}{\partial \eta}\Psi_{a}(\vk)+\Omega_{a}^{\ b}\Psi_{b}(\vk)=
\gamma_{a}^{\ bc}(\vk,\vk_{1},\vk_{2})\Psi_{b}(\vk_{1})\Psi_{c}(\vk_{2}) \label{EOM};
\]
the nonvanishing matrix elements of  $\Omega_{a}^{\ b}$ are given by
\[
\begin{split}
\Omega_{(2\bb-1)}^{\ (2\bb)}&=-1\;, \\
\Omega_{(2\bb)}^{\ (2\bb)}&=\frac{3}{2}\frac{\Omega_{\rm m}}{f_{+}^{2}}-1\;, \\
\Omega_{(2\bb)}^{\ (2\bc-1)}&=-\frac{3}{2}\frac{\Omega_{\rm m}}{f_{+}^{2}}\fw_{J} \;,
\end{split}
\]
where we denote by $\fw_\bb \equiv \Omega_\bb /\Omega_{\rm m}$ the relative fraction of the fluid $\bb$.

The nonvanishing elements of the coupling matrix $\gamma_{a}^{\ bc}$ are
\[
\begin{split}
\gamma_{(2p-1)}^{\ (2p-1)\,(2p)}(\vk,\vk_{1},\vk_{2})&=\frac{\alpha(\vk_{2},\vk_{1})}{2}\;, \\
\gamma_{(2p-1)}^{\ (2p)\,(2p-1)}(\vk,\vk_{1},\vk_{2})&=\frac{\alpha(\vk_{1},\vk_{2})}{2}\;, \\
\gamma_{(2p)}^{\ (2p)\,(2p)}(\vk,\vk_{1},\vk_{2})&=\beta(\vk_{1},\vk_{2}) \;,
\end{split}
\]
for any integer $p$. Note that there are no explicit couplings between different species in the $\gamma_{a}^{\ bc}$ matrices.

The isodensity modes are obtained under the constraint that the total density contrast vanishes, i.e.
\[
\label{constraint_iso}
\deltam =\sum_{\bb}\ \fw_{\bb}\,\delta_{\bb}=0 \;.
\]
When we consider only 2 fluids, there are 2 such modes, one decaying and one constant in time. In the following we denote by ``$\plus$'' and ``$\minus$''  the growing and decaying adiabatic modes, respectively, and by ``$\iis$'' and ``$\cis$'' the decaying and constant isodensity modes.

Since under the constraint \eqref{constraint_iso} the evolution equations decouple, the time dependence of these modes can be easily inferred. One solution is given by
\[
\begin{split}
\rtheta_{\bb}^{(\iis)}(\eta)&\propto \exp\left[-\int^{\eta}\dd\eta'\; \left(\frac{3}{2}\frac{\Omega_{\rm m}}{f_{+}^{2}}-1\right)\right] \;, \\
\delta_{\bb}^{(\iis)}(\eta)&=\int^\eta \dd \eta'\; \rtheta_{\bb}^{(\iis)}(\eta') \;,
\end{split}
\]
with
\[
\sum_\bb \fw_\bb \rtheta_{\bb}^{(\iis)}=0 \;, \label{Theta_sum_i}
\]
which automatically ensures Eq.~\eqref{constraint_iso}.
Note that, because  $\Omega_{\rm m}/f_{\plus}^2$ departs little from the value taken in an EdS cosmology, i.e.~$\Omega_{\rm m}/f_{+}^2=1$, the isodensity modes are expected to depart very weakly from
\[
\rtheta_{\bb}^{(\iis)}(\eta)\propto \exp(-\eta/2)\; ,\qquad \delta_{\bb}^{(\iis)}(\eta)=-2\rtheta_{\bb}^{(\iis)}(\eta) \;.\label{i_mode}
\]
A second set of isodensity modes is given by
\[
\rtheta_{\bb}^{(\ci)}(\eta)=0\; ,\qquad \delta_{\bb}^{(\ci)}(\eta)=\hbox{Constant}\;, \label{ci_mode}
\]
under the condition that Eq.~\eqref{constraint_iso} is satisfied.

To be specific, let us concentrate now on the case of two fluids and assume an EdS background.
In this case, the growing and decaying solutions are then proportional, respectively, to
\[
\begin{split}
u_{a}^{(\plus)} & =\left(1,1,1,1\right)^T \;,\\
u_{a}^{(\minus)} & =\left(1,-3/2,1,-3/2\right)^T \;.
\end{split}
\]
Moreover, the isodensity modes are proportional to
\be
\begin{split}
u_{a}^{(\iis)} & =\left(-2 \fw_2 ,\fw_2,2\fw_1,-\fw_1\right)^T \;, \\
u_{a}^{(\ci)} & =\left(\fw_2 ,0,-\fw_1,0\right)^T \;.
\end{split}
\ee

We are then in the position to write down the linear propagator $g_{a}^{\ b}(\eta,\eta_0)$, satisfying
\[
\frac{\partial}{\partial \eta}g_{a}^{\ b}(\eta,\eta_0)+\Omega_{a}^{\ c} (\eta) g_{c}^{\ b}(\eta,\eta_0)=0 \;,
\]
with
\[
g_{a}^{\ b}(\eta,\eta)=\delta_{a}^{\ b}.
\]
For two fluids and an EdS background, its explicit form has been
given in \cite{2010PhRvD..81b3524S} and reads
\begin{widetext}
\[
\begin{split}
g_{a}^{\ b}(\eta,\eta_0)= \ &\frac{e^{\eta-\eta_0}}{5}
\left(
\begin{array}{cccc}
 3 \fw_1 & 2 \fw_1 & 3 \fw_2 & 2 \fw_2 \\
 3 \fw_1 & 2 \fw_1 & 3 \fw_2 & 2 \fw_2 \\
 3 \fw_1 & 2 \fw_1 & 3 \fw_2 & 2 \fw_2 \\
 3 \fw_1 & 2 \fw_1 & 3 \fw_2 & 2 \fw_2
\end{array}
\right)
+\frac{e^{-\frac32(\eta-\eta_0)}}{5}
\left(
\begin{array}{cccc}
 2 \fw_1 & -2 \fw_1 & 2 \fw_2 & -2 \fw_2 \\
 -3 \fw_1 & 3 \fw_1 & -3 \fw_2 & 3 \fw_2 \\
 2 \fw_1 & -2 \fw_1 & 2 \fw_2 & -2 \fw_2 \\
 -3 \fw_1 & 3 \fw_1 & -3 \fw_2 & 3 \fw_2
\end{array}
\right) \\
&+
e^{-\frac12(\eta-\eta_0)}
\left(
\begin{array}{cccc}
 0 & -2 \fw_2 & 0 & 2 \fw_2 \\
 0 & \fw_2 & 0 & -\fw_2 \\
 0 & 2 \fw_1 & 0 & -2 \fw_1 \\
 0 & -\fw_1 & 0 & \fw_1
\end{array}
\right)
+
\left(
\begin{array}{cccc}
 \fw_2 & 2 \fw_2 & -\fw_2 & -2 \fw_2 \\
 0 & 0 & 0 & 0 \\
 -\fw_1 & -2 \fw_1 & \fw_1 & 2 \fw_1 \\
 0 & 0 & 0 & 0
\end{array}
\right) \;. \label{linear_g}
\end{split}
\]
\end{widetext}
In the following we explore how this propagator is changed in the eikonal approximation.

Armed with the explicit form of the propagator, the equation of motion \eqref{EOM} can then be written in integral form as
\[
\begin{split}
\Psi_{a}(\vk,\eta)=g_{a}^{\ b}(\eta,\eta_{0})\Psi_{b}(\vk,\eta_{0})+\\
\int_{\eta_{0}}^{\eta}\dd\eta'
g_{a}^{\ b}(\eta,\eta')\gamma_{b}^{\ cd}(\vk_{1},\vk_{2})\Psi_{c}(\vk_{1},\eta')\Psi_{d}(\vk_{2},\eta')\label{EOMint_0} \;.
\end{split}
\]
It is this form that can be used in practice for PT calculations.

\subsection{Basis redefinition}

In the presence of multiple degrees of freedom, it can be technically judicious to rotate the basis representation of the multiplet into the eigenvectors of the propagators. This makes PT calculations less cumbersome because in this representation components mix together only at the vertex position and not along the linear evolution.
Mathematically, this amounts to make the multiplet transformation
\[
\Psi_{\alpha} \equiv \rcomp_{\alpha}^{\ a}\,\Psi_{a} \;, \label{proj}
\]
where ${\cal R}$ explicitly reads
\[
\rcomp_{\alpha}^{\ a}=\left(
\begin{array}{cccc}
 \frac{3 \fw_1}{5} & \frac{2 \fw_1}{5} & \frac{3 \fw_2}{5}  &
   \frac{2 \fw_2}{5}  \\
 \frac{2 \fw_1}{5} & -\frac{2 \fw_1}{5} & \frac{2 \fw_2}{5}  &
   -\frac{2 \fw_2}{5}  \\
 0 & 1 & 0 & -1 \\
 1 & 2 & -1 & -2
\end{array}
\right),
\]
in such a way that each component of $\Psi_{\alpha}$ now represents an eigenvector of the propagator.
Note that this relation can be inverted into
\[
\Psi_{a}=\ircomp_{a}^{\ \alpha}\,\Psi_{\alpha} \;,
\]
where ${\cal S}$ is the inverse of ${\cal R}$, satisfying
\[
\ircomp_{a}^{\ \alpha}\,\rcomp_{\alpha}^{\ b}=\delta_{a}^{\ b}\;.
\]

In this representation Eq.~(\ref{EOMint_0}) becomes
\[
\begin{split}
\Psi_{\alpha}(\vk,\eta)=g_{\alpha}^{\ \beta}(\eta,\eta_{0})\Psi_{\beta}(\vk,\eta_{0})+\\
\int_{\eta_{0}}^{\eta}\dd\eta'
g_{\alpha}^{\ \beta}(\eta,\eta')\gamma_{\beta}^{\ \delta\lambda}(\vk_{1},\vk_{2})\Psi_{\delta}(\vk_{1},\eta')\Psi_{\lambda}(\vk_{2},\eta')\label{EOMint} \;,
\end{split}
\]
where
\[
g_{\alpha}^{\ \beta} \equiv \rcomp_{\alpha}^{\ a}\,g_{a}^{\ b}\,\ircomp_{b}^{\ \beta} \;,
\]
and
\[
\gamma_{\alpha}^{\ \beta\delta} \equiv \rcomp_{\alpha}^{\ a}\,\gamma_{a}^{\ bd}\,\ircomp_{b}^{\ \beta}\,\ircomp_{d}^{\ \delta}\; .
\]
Note that  the linear propagator takes a diagonal form,
\[
g_{\alpha}^{\ \beta}(\eta,\eta_{0})=\left(
\begin{array}{cccc}
e^{\eta-\eta_{0}}&0&0&0\\
0&e^{-3/2(\eta-\eta_{0})}&0&0\\
0&0&e^{-1/2(\eta-\eta_{0})}&0\\
0&0&0&1
\end{array}
\right)\;.
\]

The vertex matrix has however a more complex structure than the original one. It still contains a significant number of zero components:
\be
\begin{split}
\gamma_{\gamma}^{\ \alpha\beta}&=0\ \ \hbox{for\ \ }\gamma\ \hbox{and}\ \ \alpha=\plus \;,\ \minus\ \ \hbox{and}\ \ \beta= \iis\;,\ \cis\;;\\
\gamma_{\gamma}^{\ \cis\cis}&=0\;;\\
\gamma_{\gamma}^{\ \alpha\beta}&=0\ \ \hbox{for\ \ }\gamma=\iis \;,\ \cis\ \ \ \hbox{and}\ \ \alpha \    \hbox{and}\  \beta= \plus \;,\ \minus \;;\\
\gamma_{\iis}^{\ \alpha\beta}&=0\ \ \hbox{for\ \ }\alpha\ \hbox{or}\ \beta=\cis\;. \label{table_gamma}
\end{split}
\ee
At this stage, it is worth mentioning a couple of properties regarding the interplay between the modes. Isodensity modes can induce adiabatic modes only when two of them interact together (first line of Eq.~\eqref{table_gamma}); moreover,
the constant isodensity mode does not induce adiabatic modes at any order if there is no \textsl{decaying} isodensity mode (second line of Eq.~\eqref{table_gamma}).

With this formalism it is then easy to trace back the impact of the various modes. When the greek indices are restricted to $\plus$ and $\minus$, calculations are done for the adiabatic modes only. Then the dynamics is the same as the one of a single fluid. The impact of the existence of multiple components is then obtained from the effects of the third and fourth modes.

\subsection{Diagrammatic representation and exact 1-loop calculation}

\begin{figure}
\centerline{\epsfig {figure=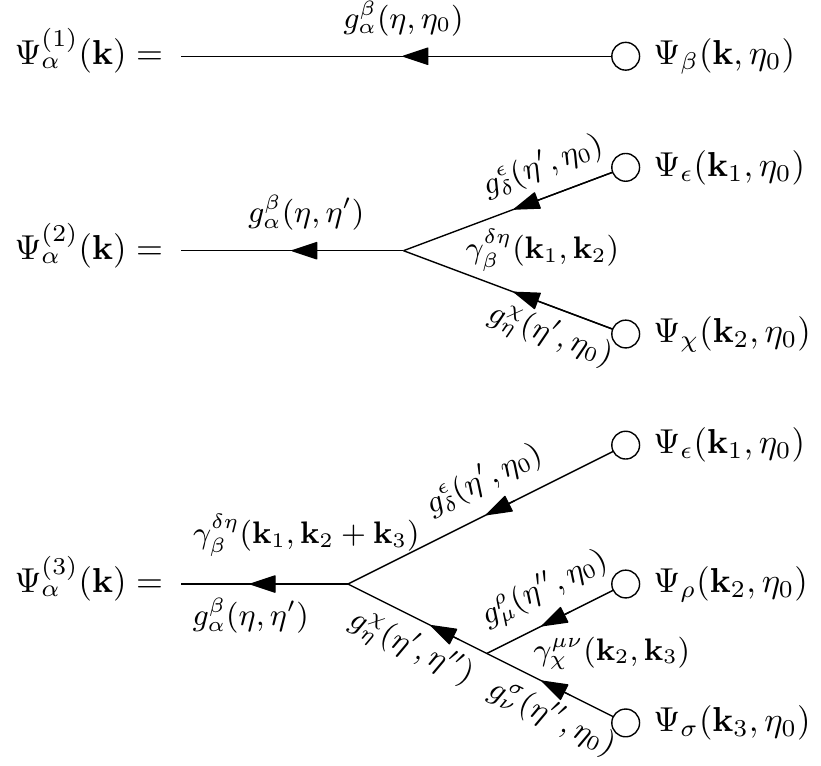,width=8cm}}
\caption{Diagrammatic representation of the series expansion of $\Psi_{\alpha}(\vk)$ up to second order in the initial conditions at $\eta=\eta_0$. Time increases along each segment according to the arrow and each segment bears a factor $g_{\alpha}^{\ \beta}(\eta,\eta_0)$ if $\eta_{0}$ is the initial time and $\eta$ is the final time.
At each initial point and each vertex point there is a sum over the component indices; a sum over the incoming wave modes is also implicit and, finally, the time coordinate of the vertex points is integrated from $\eta=\eta_0$ to the final time $\eta$ according to the time ordering of each diagram. }
\label{MultiExpansPsi}
\end{figure}

The integral representation of the solution to the field equations, Eq.~\eqref{EOMint}, can be  graphically represented in terms of diagrams (see for instance \cite{Crocce:2005xy} for details). If we write the solution in the form of a series expansion in the number of initial fields,
\be
\Psi_\alpha (\vk,\eta) = \sum_{n=1}^\infty \Psi^{(n)}_\alpha(\vk,\eta)\;,
\ee
(here we are using the propagator basis representation) to compute the 1-loop power spectrum we just need to expand up to third order. Formally, we have
\begin{align}
\Psi_{\alpha}^{(1)}(\vk,\eta)=&\ g_\alpha^{\ \beta} (\eta,\eta_0) \Psi_{\beta}(\vk,\eta_{0}) \;, \\
\Psi_{\alpha}^{(2)}(\vk,\eta)=&\ \Gamma_{\alpha}^{\ \beta\delta}(\vk, \vk_{1},\vk_{2};\eta,\eta_{0}) \Psi_{\beta}(\vk_{1},\eta_{0})\,
\Psi_{\delta}(\vk_{2},\eta_{0})\;, \\
\Psi_{\alpha}^{(3)}(\vk,\eta)=&\ \Gamma_{\alpha}^{\ \beta\delta\sigma}(\vk, \vk_{1},\vk_{2},\vk_{3};\eta,\eta_{0})
\nonumber\\
&\times\Psi_{\beta}(\vk_{1},\eta_{0})
\, \Psi_{\delta}(\vk_{2},\eta_{0})
\,\Psi_{\sigma}(\vk_{3},\eta_{0})\;,
\end{align}
where on the right-hand side convolutions with the wave modes are implicit.
The functions $\Gamma_{\alpha}^{\ \beta\delta}$  and $\Gamma_{\alpha}^{\ \beta\delta\sigma}$ are simply the
2 and 3-point propagators introduced in \cite{2008PhRvD..78j3521B}, taken at tree order. Their explicit expressions are given in the Appendix.
They are defined in such a way that they are symmetric under simultaneous permutations of upper indices and their corresponding wave modes, e.g.~$\Gamma_{\alpha}^{\ \beta\delta}(\vk, \vk_{1},\vk_{2};\eta,\eta_{0}) = \Gamma_{\alpha}^{\ \delta\beta}(\vk, \vk_{2},\vk_{1};\eta,\eta_{0})$. The diagrammatic representation of these three equations is shown in Fig.~\ref{MultiExpansPsi}. 
\begin{figure}
\centerline{\epsfig {figure=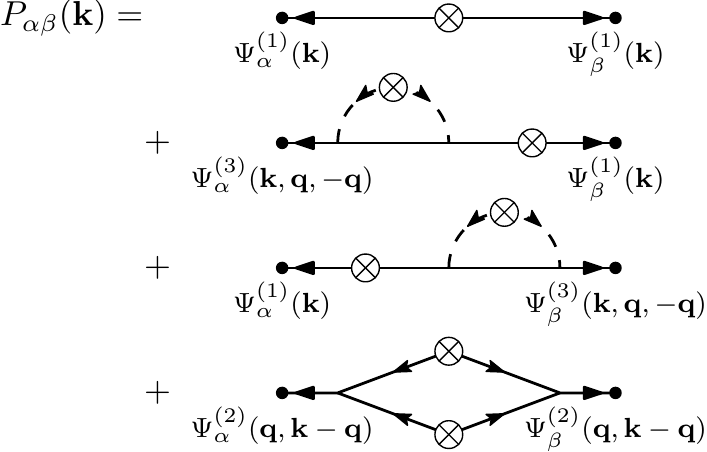,width=8cm}}
\caption{Diagrammatic representation of the power spectra up to 1-loop order.  The symbol $\otimes$ represents the power spectra at initial
time. Note that the power spectra and cross spectra of all 4 modes should be taken into account.}
\label{MultiOneLoop}
\end{figure}

Assuming Gaussian initial conditions and initial power spectra $P^{\ini}_{\alpha \alpha'}(k)$ defined by
\be
\langle \Psi_{\alpha}(\vk,\eta_0) \Psi_{\alpha}(\vk',\eta_0) \rangle  = \delta_{\rm D}(\vk + \vk') P^{\ini}_{\alpha \alpha'}(k)\;,
\ee
the tree-level power spectra are linearly related to the initial ones by
\[
P^{(0)}_{\alpha\alpha'}(k, \eta)=  g_\alpha^{\ \beta}(\eta,\eta_0) g_{\alpha'}^{\ \beta'}(\eta,\eta_0) P^{\ini}_{\beta \beta'}(k) \;.
\]
The graphical  representation of this relation is shown in the first diagram of Fig.~\ref{MultiOneLoop}.
The 1-loop contribution is instead given by
\[
\begin{split}
&P^{(1)}_{\alpha\alpha'}(k,\eta)=   6\,
\Gamma_{\alpha}^{\ \beta\delta\sigma}(\vk, \vk,\vq,-\vq;\eta,\eta_{0}) \, g_{\alpha'}^{\ \beta'}(\eta,\eta_{0})\\
&\qquad\qquad\qquad\qquad \times P^{\ini}_{\delta\sigma}(q)
P_{\beta\beta'}^{\ini}(k) \\
&\qquad + 2\,\Gamma_{\alpha}^{\ \beta\delta}(\vk,\vk-\vq, \vq;\eta,\eta_{0}) \Gamma_{\alpha'}^{\ \beta'\delta'}(\vk,\vk-\vq,\vq;\eta,\eta_{0})\\
&\qquad\qquad\qquad\qquad \times \,P^{\ini}_{\beta\beta'}(|\vk-\vq|)
\,P^{\ini}_{\delta\delta'}(q) \;,
\end{split}
\label{1loopcorr}
\]
when all symmetry factors are properly taken into account. The first two lines of this expression represent the 1-loop corrections to the (2-point) propagator (second and third diagrams of Fig.~\ref{MultiOneLoop}), while the last two lines  contain the irreducible 1-loop contribution (last diagram of Fig.~\ref{MultiOneLoop}).
Note that, in this expression, all modes---adiabatic and isodensity---are taken into account when necessary.  This is at variance with what is usually done in PT calculations, where only the growing adiabatic mode is included in the loops and in the initial conditions.

In practice, for the initial conditions we use the initial transfer functions
obtained from the code CAMB~\cite{Lewis:1999bs}, with concordance model  cosmological parameters. In this context, we assume the primordial metric perturbations to be adiabatic and the initial conditions to obey Gaussian statistics. More precisely, the initial power spectra can be written as
\[
P^{\ini}_{\alpha\beta}(k)=T_{\alpha}(k,\eta_{0})T_{\beta}(k,\eta_{0})P^{\rm prim. }(k) \;,
\]
where $T_{\alpha}(k,\eta_{0})$ are the mode transfer functions computed at time $\eta=\eta_{0}$ and $P^{\rm prim. }(k)$ is the power spectrum of the primordial curvature perturbations. More details on the choice of the initial conditions and a description of the shape of the transfer functions can be found in \cite{2012PhRvD..85f3509B}.

The initial conditions have been chosen at $z_0=980$, at a time when the baryons decouple from the photon plasma and free fall into the
dark matter potentials. Note that, starting at this very high redshift, scales that are of interest today might have been comparable to the Hubble radius at the initial time. On these scales, General Relativity substantially deviates from the purely Newtonian description used in PT. Moreover, the transfer functions generated by CAMB are given in a specific gauge and  choosing a different one may lead to different initial conditions. For this reason the choice of the initial conditions has to be taken with a little care.
Fortunately, as explained in \cite{2012PhRvD..85f3509B} (see also \cite{Chisari:2011iq}), there exists a choice of variables for which at linear order the relativistic equations exactly reduce to the Newtonian ones even in the case of multiple pressureless fluids; in this case the matching between the transfer function generated by CAMB and the PT evolution can be consistently performed.

We thus have all the necessary ingredients to compute the density power spectra up to 1-loop, which we do numerically. We now focus on the results obtained at high $z$, where the impact of the isodensity modes are clearly visible.

\begin{figure}
\centerline{\epsfig {figure=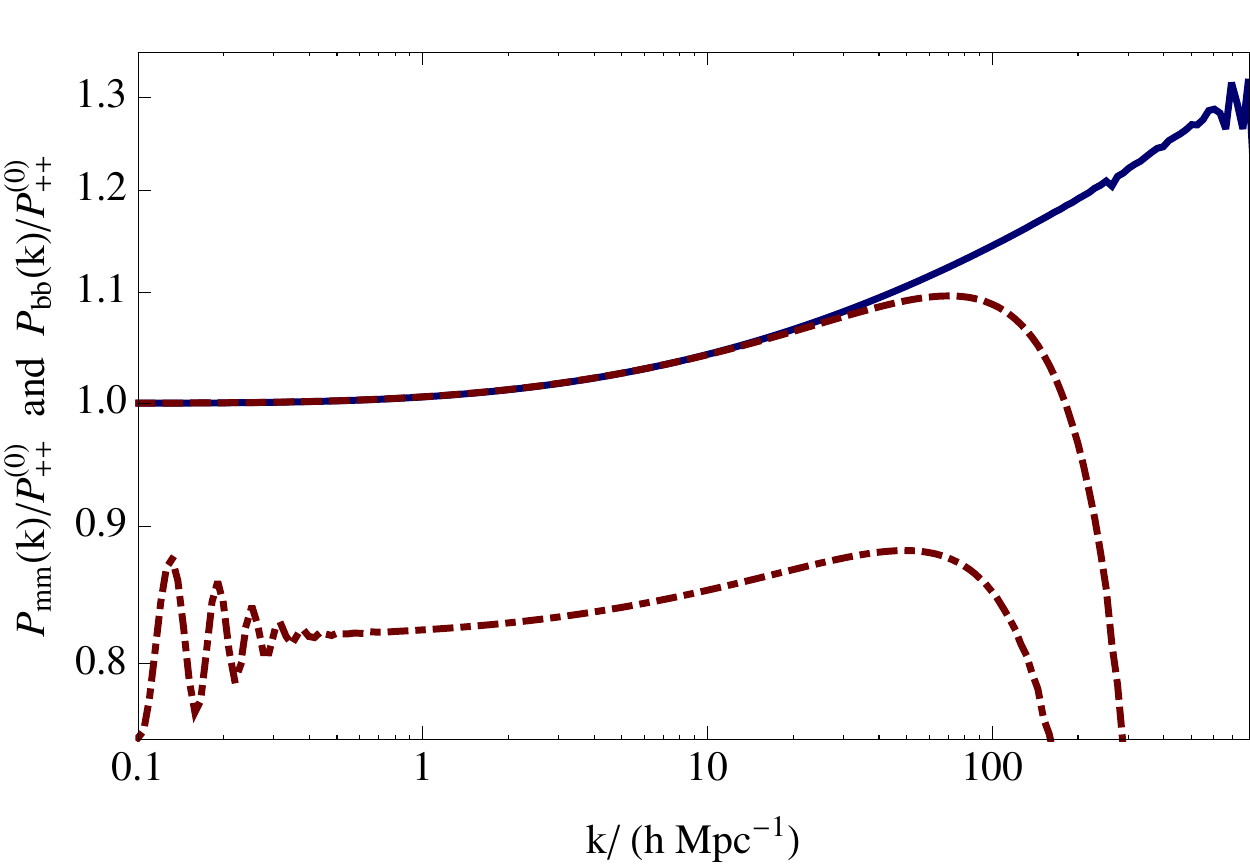,width=8cm}}
\caption{Power spectrum of the total matter fluctuations (solid blue and dashed red line) and of baryonic fluctuations (dot-dashed red  line) 
computed at 1-loop order as a function of $k$ at $z=40$, in units of the \textsl{linear} total matter power spectrum  $P_{++}^{(0)}$ (i.e.~of the adiabatic mode). For the total matter fluctuations, the power spectrum is computed either including the contribution of the adiabatic modes only (blue line) and taking into account all modes including also the isodensity modes (dashed red line). In the two red lines one can see the damping of modes due to the isodensity modes. 
}
\label{P1loopz40}
\end{figure}

\begin{figure}
\centerline{\epsfig {figure=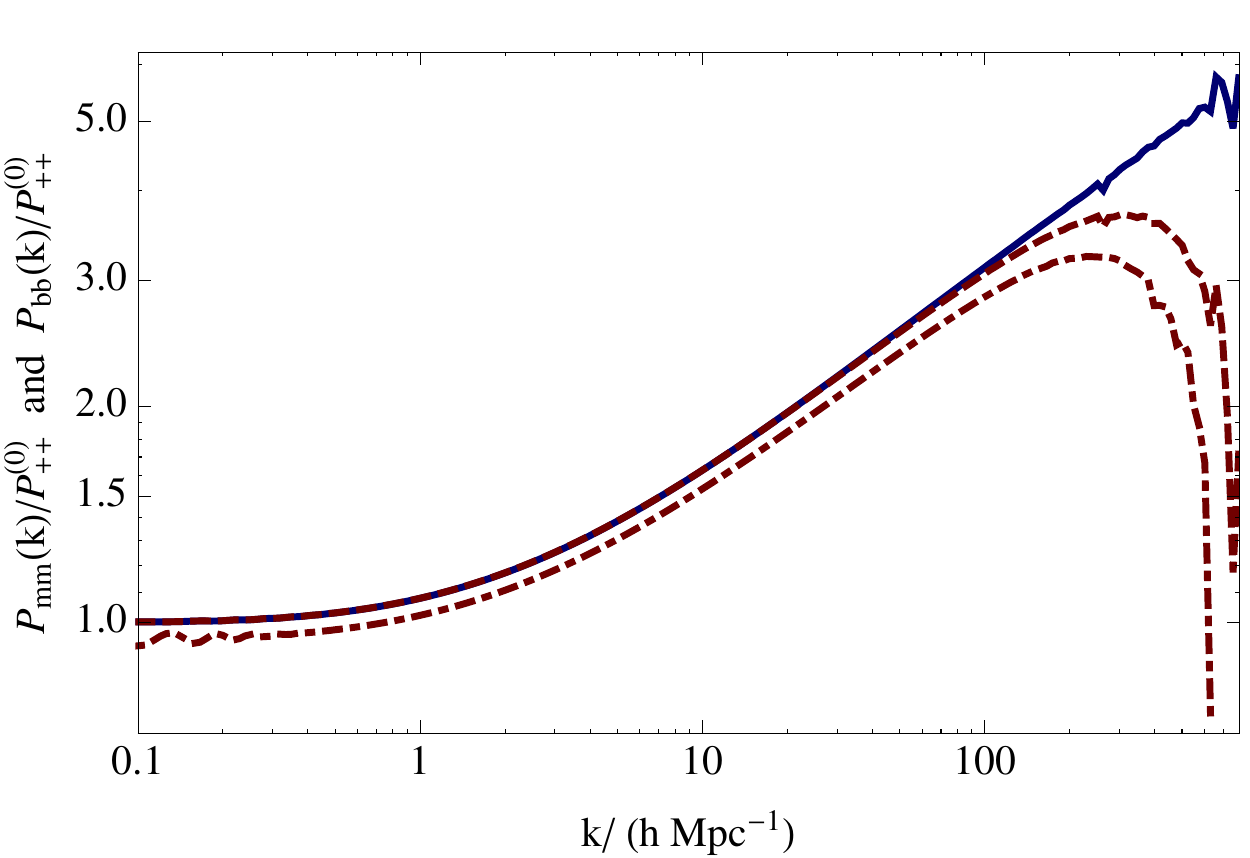,width=8cm}}
\caption{The same as in Fig.~\ref{P1loopz40}, but at $z=10$.}
\label{P1loopz10}
\end{figure}

The resulting 1-loop power spectrum of the total density contrast,  $P_{\rm m m} = P^{(0)}_{\rm m m} + P_{\rm m m}^{(1)}$, and of the density contrast in the baryonic component 
are shown for $z=40$ in Fig.~\ref{P1loopz40} and for $z=10$ in Fig.~\ref{P1loopz10}. In these figures, one can clearly see that the baryons are still significantly less clustered than the overall matter field. In other words, some subleading modes---such as the decaying isodensity modes---do play a significant role, even for the linear theory, at these redshifts. The plots also exhibit the rise of the amplitude of the power spectrum due to 1-loop effects. For scales below $k=50\,h/\Mpc$ this is expected and most of the effect  is due to the coupling to the adiabatic modes. However, for larger $k$ the 1-loop power spectrum is damped; as explained below, this is entirely due to the decaying isodensity modes. Note that at very high $k$ the numerical integration exhibits numerical noise, clearly visible on the figures,  due to the cancellation of large terms.

At lower redshifts, $z<10$, the effect of the isodensity modes, although still present, is nearly washed out. In this regime the 1-loop corrections due to the adiabatic modes represent the dominant contributions.

\subsection{Leading $k$ behavior}
\label{sec:c}

In this subsection we examine the expected behavior of the 1-loop correction to the power spectra at large $k$.
The aim is to pin down the origin of the  damping of the 1-loop correction of the density power spectrum shown in the previous subsection.
To do this, we can view  the 1-loop correction as a functional form of the linear spectra or, more particularly, of the \textsl{moments} of the power spectra, defined as
\[
M^{(p)}_{\alpha\beta} \equiv {4\pi}\int\dd q\ q^{2+p} P^{\ini}_{\alpha\beta}(q).\label{Mpalphabetadef}
\]
The idea is then to  organize  the 1-loop correction in its dependence on such moments, starting from $p=-2$.

In the high-$k$ limit, $k/q \gg 1$, both vertices  \eqref{alpha} and \eqref{beta} behave as $\vk \cdot \vq /q^2$, each of them effectively introducing a power of $k/q$.
Given that the 1-loop  contribution to the  power spectra contains products of two vertices integrated over the (3-d)  internal momentum $\vq$, barring accidental cancellations one expects  the leading $k$ contribution  to be of the form $k^{2}\sigma^{2}_{\alpha\beta}$ with
\[
\sigma^{2}_{\alpha\beta}\equiv\frac{4\pi}{3}\int\dd q \; P^{\ini}_{\alpha\beta}(q)=\frac{1}{3}M^{(-2)}_{\alpha\beta}.\label{sigmaalphabetadef}
\]

The explicit calculation can be done exactly and, at leading order in $k/q$, the most growing contribution to the 1-loop correction to the power spectrum reads
\[
\begin{split}
P^{(1)}_{\plus\plus}(k, \eta) =
-\frac{8\,\fw_{1}\,\fw_{2}\,e^{2(\eta-\eta_{0})}}{5}\ k^{2}\sigma_{\iis\iis}^{2}
\left[P_{\plus\plus}^{\ini}(k)\right.\\
\left.
-\frac{1}{14}P_{\plus\minus}^{\ini}(k)-\frac{\fw_{2}-\fw_{1}}{5}P_{\plus\iis}^{\ini}(k)+\frac{\fw_{2}-\fw_{1}}{4}P_{\plus\cis}^{\ini}(k)
\right.\\
\left.
-\frac{\fw_{1}\fw_{2}}{10}\left(P_{\iis\iis}^{\ini}(k)+P_{\cis\cis}^{\ini}(k)-2P_{\iis\cis}^{\ini}(k)\right)
\right].
\label{onelooplk}
\end{split}
\]
Interestingly, the whole correction is negative---which confirms the behavior observed in Figs.~\ref{P1loopz40} and \ref{P1loopz10}---and, most importantly, only proportional to $\sigma_{\iis\iis}^{2}$. 
In other words,
\textsl{when the decaying isodensity mode is absent in the initial conditions, there is
no  corrective term at this order in $k/q$}---the next correction is proportional to $M^{(0)}_{\alpha\beta}$. We can thus attribute the decaying behavior of the 1-loop power spectrum in Figs.~\ref{P1loopz40} and \ref{P1loopz10} at $k>50\,h/\Mpc$ exclusively to the decaying isodensity initial conditions.

That only $\sigma_{\iis\iis}^{2}$ appears in the loop correction is somewhat surprising but can be explained by the fact that, even though the adiabatic modes intervene in the loops, their contribution at lowest order in $q/k$  sum up to zero in the final spectra \footnote{Note that the constant isodensity mode does not contribute to loops simply because $\gamma_{\alpha}^{\ \cis \beta} (\vk, \vq, \vk) =  0$ at lowest order in $q/k$.}. In particular, this cancellation is due to the fact that the vertices with one leg being a soft adiabatic mode are diagonal at lowest order in $q/k$. Indeed, at large $k$ they read
\begin{align}
\gamma_{\alpha}^{\ + \beta} (\vk, \vq, \vk) &=  \frac{\vq \cdot \vk}{2 q^2} \; \delta_\alpha^{\ \beta} \;, \label{vert_+}\\
\gamma_{\alpha}^{\ - \beta} (\vk, \vq, \vk) &=  - \frac{3\vq \cdot \vk}{2 q^2} \; \delta_\alpha^{\ \beta} \label{vert_-} \;.
\end{align}

\begin{figure*}
\centerline{\epsfig {figure=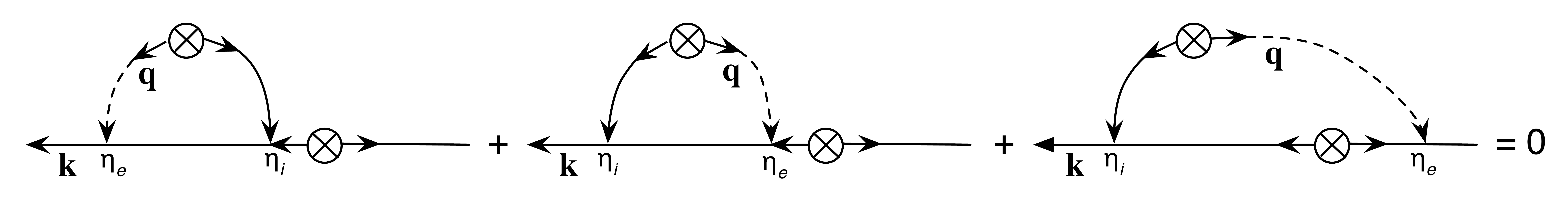,width=17cm}}
\caption{This plot illustrates the central property of diagram complementarity in the leading $k$ regime. The sum of these 3 diagrams vanishes
when the dashed line corresponds to a given \textsl{adiabatic} mode and when the vertex it is connected to is computed in the soft limit $q/k \to 0$, Eqs.~\eqref{vert_+} and \eqref{vert_-}, or in the eikonal limit, Eq.~(\ref{eikovertex}).
The only case for which these 3 contributions do not cancel is when the intervening mode is in the decaying isodensity mode. }
\label{AdiabEikos}
\end{figure*}

To see how the cancellation works, let us consider the diagrams of Fig.~\ref{AdiabEikos} and take, for instance, the dashed line belonging to the loops in the growing adiabatic mode; its  vertex is thus given by Eq.~\eqref{vert_+}. Let us denote by $\eta_e$ the time of interaction involving the dashed line and by $\eta_i$ the other time of interaction. These three diagrams contribute to the 1-loop correction to the power spectrum by something proportional to 
\begin{widetext}
\be
\tilde P_{\alpha \alpha'} (k)= \int_{\eta_0}^{\eta}\dd \eta_i  \int \dd^3 \vq \left[ \int_{\eta_i}^{\eta} \dd \eta_e  \gamma_\alpha^{\ + \beta} \gamma_{\beta}^{\ c\delta} P^{\ini}_{\delta \alpha'}(k)
+ \int_{\eta_0}^{\eta_i} \dd \eta_e  \gamma_\beta^{\ + \delta} \gamma_{\alpha}^{\ \gamma \beta}  P^{\ini}_{\delta \alpha'}(k)  + \int_{\eta_0}^\eta \dd \eta_e \gamma_\alpha^{\ \gamma \beta} \gamma_{\alpha'}^{\ + \beta'}  P_{ \beta \beta'}^{\ini}(k)\right] P_{+\gamma}(q; \eta_e,\eta_i) \;,
\label{eqc}
\ee
\end{widetext}
where $P_{+ \gamma}$ is the unequal time power spectrum of the soft modes taken at times $\eta_{i}$ and
$\eta_{e}$ and we have not included the time dependence along the principal line, which is the same for the three diagrams.  The sum of these three contributions vanishes by using Eq.~\eqref{vert_+} and the fact that the momentum of the last contribution is equal but of opposite sign to the one of the first two contributions. Thus, the first two diagrams of Fig.~\ref{AdiabEikos} exactly cancel with the last diagram of the same figure. The same holds for the decaying adiabatic mode.
We will come back to this property---and generalize it to any loop order---in the next section, in the context of the eikonal approximation.

It is interesting to note that this correction involves a whole set of initial power spectra, not only the adiabatic growing mode,
although only the most growing contribution has been selected in the resulting expression in Eq.~(\ref{onelooplk}).
Note also that the leading time dependence of these terms is the same as the linear adiabatic growing mode.
We are then led to the following effective result: at large enough $k$ all the initial power spectra are proportional and then the 1-loop power spectrum can be written as,
\[
P_{\plus\plus}(k,\eta)\approx e^{2(\eta-\eta_{0})}P^{\ini}_{\plus\plus}(k)\left[1-\left(\frac{k}{ k_{\rm damp}}\right)^{2}\right] \;,
\label{kdampval2}
\]
where the scale $k_{\rm damp}$ depends on the cosmological model. 
For the concordant $\Lambda$CDM model we find
\begin{equation}
k_{\rm damp}\approx 380\ h {\rm Mpc}^{-1}\;\label{kdampval}.
\end{equation}

In this form, the large-scale modes are only partially taken into account. The $k$ dependence of the corrective factor is that
of the 1-loop corrective term.
The eikonal approach tells us precisely  how these early time leading contributions can be resumed nonperturbatively, i.e.~what is the full $k$ dependence of the damping factor, not only its 1-loop contribution as in the previous expression. Thus, we now turn to the use of the eikonal approximation.

\section{The eikonal approximation}
\label{sec:SectEikonal}

The eikonal approximation has been formally introduced in \cite{2012PhRvD..85f3509B}. It is based on a separation of scales: the effect of the large-scale modes is treated as a change of the background for the small-scale modes, explicitly operating as a redefinition of the coefficients in the dynamical equations. In the following we will review the construction of the background dependent dynamical system in the eikonal limit. We will then solve it in the linear regime incorporating the impact of the long-wavelength modes \textsl{in a nonperturbative way}. We will focus on the computation of the power spectra, particularly on the impact of the baryonic-CDM isodensity modes on their small-scale behavior.

\subsection{Equations of motion}

The eikonal approximation consists in splitting the right-hand side  of Eq.~(\ref{EOM}) into two different contributions:
one coming from the coupling of two modes of very different amplitudes, i.e.~$k_{1}\ll k_{2}$ or $k_{2}\ll k_{1}$, and  one coming from the coupling of two modes of  comparable amplitudes. By doing so the equations of motion \eqref{EOM} can then be rewritten as
\[
\begin{split}
&\frac{\partial}{\partial \eta}\Psi_{a}(\vk)+\Omega_{a}^{\ b}\Psi_{b}(\vk)-\Xi_{a}^{\ b}(\vk)\Psi_{b}(\vk) \\
&=
\left[ \gamma_{a}^{\ bc}(\vk,\vk_{1},\vk_{2})\Psi_{b}(\vk_{1})\Psi_{c}(\vk_{2}) \right]_{\rm \cal H}\label{eikonalNL} \;,
\end{split}
\]
with
\[
\Xi_{a}^{\ b}(\vk,\eta)\equiv  \int_{{\cal S}}\dd^{3}\vq \; \left[\egamma_{a}^{\ cb}(\vk,\vq,\vk)+\egamma_{a}^{\ bc}(\vk,\vk,\vq)\right]\Psi_{c}(\vq,\eta)  \;. \label{Xi_def}
\]
The key point is that in Eq.~(\ref{Xi_def}) the domain of integration is restricted to the soft momenta---hence the ${\cal S}$ index---for which $q\ll k$. Conversely, on the right-hand side of Eq.~\eqref{eikonalNL} the convolution is done excluding the soft domain, i.e.~it is over hard modes or modes of comparable size---hence the ${\cal H}$ index for the integral domain in the convolution.

The coefficient $\Xi_{a}^{\ b}(\vk)$ in the evolution equation (\ref{EOM}) depends on the initial conditions of the soft modes only.
In the limit of separation of scales, $q\ll k$, $\Xi_{a}^{\ b}(\vk)$ can be treated as a random quantity in this equation.
Moreover, using Eqs.~\eqref{alpha} and \eqref{beta},  the leading expression of the coupling matrix in the limit $q\ll k$ is obtained for $\alpha(\vq,\vk)\approx (\vq \cdot \vk)/q^{2}$, $\alpha(\vk,\vq)\approx0$,
$\beta(\vq,\vk) =\beta(\vk,\vq)\approx (\vq \cdot \vk)/(2q^{2})$.

In the case of two fluids, $\Xi_{a}^{\ b}$ is given by the sum of an adiabatic and an isodensity contribution,
\[
\Xi_{a}^{\ b}(\vk,\eta)=\Xi^{({\rm ad})} (\vk,\eta)  \; \delta_{a}^{\ b}+{\Xi^{(\iis)}\!{}_{a}^{\ b}}(\vk,\eta)\;.
\label{Xi_3}
\]
Here
\[
\Xi^{({\rm ad})}(\vk,\eta) \equiv \int_{\cal S} \dd^3 \vq \frac{\vk \cdot \vq }{q^2} \big( \Theta^{(+)} (\vq,\eta) + \Theta^{(-)} (\vq,\eta)\big) \;,
\]
where $\Theta^{(+)}$ and $\Theta^{(-)}$ are, respectively, the growing and decaying adiabatic modes of the long-wavelength velocity field.
The isodensity contribution $\Xi^{(\iis)}\!{}_{a}^{\ b}$ contains the decaying isodensity mode given in Eq.~\eqref{i_mode}. As  $\Theta_\bb^{(\iis)}$ satisfies Eq.~\eqref{Theta_sum_i},  the matrix $\Xi^{(\iis)}\!{}_{a}^{\ b}$ cannot be proportional to the identity, but it can be written as
\[
\Xi^{(\iis)}\!{}_{a}^{\ b} = \Xi^{(\iis)} \; h_{a}^{\ b} \;, \quad h_{a}^{\ b} \equiv
\left(
\begin{array}{cccc}
 \fw_2 & 0 & 0 & 0 \\
 0 & \fw_2 & 0 &  0 \\
 0 & 0 & -\fw_1 & 0 \\
 0 & 0 & 0 & - \fw_1
\end{array}
\right) \;, \label{iso_prop}
\]
with
\[
\Xi^{(\iis)} \equiv \frac{1}{\fw_2} \int_{\cal S} \dd^3 \vq \frac{\vk \cdot \vq }{q^2} \Theta_1^{(\iis)}  = - \frac{1}{\fw_1} \int_{\cal S} \dd^3 \vq \frac{\vk \cdot \vq }{q^2} \Theta_2^{(\iis)}  \;.
\label{XiisDef}
\]
Using the projection \eqref{proj} this definition can also be rewritten as 
\[
\Xi^{(\iis)} = \int_{\cal S} \dd^3 \vq \frac{\vk \cdot \vq }{q^2} \Psi_{\iis}  \;.
\label{XiisDef2}
\]
Note that, because of Eq.~\eqref{ci_mode}, the constant isodensity mode does not contribute
to $\Xi_{a}^{\ b}$.

\subsection{Propagators and power spectra}

We are now interested in computing the resummed propagators in the eikonal approximation, under the modulation of the long-wavelength modes in Eq.~\eqref{Xi_3}.  More specifically, we are  interested in solving Eq.~\eqref{eikonalNL} in its \textsl{linear} regime, dropping the right-hand side.

Let us define $\xi_{a}^{\ b}(\vk,\eta,\eta_{0};\Xi)$ such that
\[
\Psi_{a}(\vk,\eta;\Xi)=\xi_{a}^{\ b}(\vk,\eta,\eta_{0};\Xi)\Psi_{b}(\vk,\eta_{0})\label{psiaevol}
\]
is the linear solution of Eq.~(\ref{eikonalNL}) when its right-hand side is dropped, i.e.~in its linear approximation.
The linear propagator $\xi_{a}^{ \ b}(\vk,\eta,\eta_{0};\Xi)$ is  a function of $\Xi_a^{\ b}(\vk, \eta)$.
Thus, although it is the \textsl{linear} propagator of the eikonal system, it is a \textsl{fully nonlinear} object from the point of view of
the original system, Eq.~\eqref{EOM}.

One can then define the associated power spectra $P_{ab}(\vk, \eta;\Xi)$ as
\be
\mg \Psi_{a}(\vk,\eta;\Xi_{\vk}) \Psi_{b}(\vk',\eta;\Xi_{\vk'})\md
\equiv
\Dirac(\vk+\vk')P_{ab}(\vk,\eta;\Xi) \;,
\ee
which from (\ref{psiaevol}) read
\be
P_{ab}(\vk;\Xi)=\xi_{a}^{\ c}(\vk,\eta,\eta_{0};\Xi_{\vk})\xi_{b}^{\ d}(-\vk,\eta,\eta_{0};\Xi_{-\vk}) P^{\ini}_{cd}(k)
\ee
(assuming the initial conditions are $\Psi_{d}(\vk,\eta_{0})$ at time $\eta_{0}$).
As for the propagator defined in Eq.~\eqref{psiaevol}, the power spectra in this equation depend on a particular realization of  $\Xi$.
In order to compute the \textsl{expected} power spectra, and not just a particular realization, we need to average over the large-scale modes.
One can thus define the full ensemble average of $P_{ab}(\vk,\Xi)$ as
\be
\begin{split}
\hP_{ab}(k,\eta)=& \ \mg\xi_{a}^{\ c}(\vk,\eta,\eta_{0};\Xi_{\vk})\xi_{b}^{\ d}(-\vk,\eta,\eta_{0};\Xi_{-\vk})\md_{\Xi}\\
& \times P^{\ini}_{cd}(k) \;,
\end{split}
\label{spectraAdiab}
\ee
whose expectation value is taken over the realizations of ${\Xi(\vk)}$, i.e.~on the statistical properties of the large-scale modes.

Let us now inspect the effect of the large-scale modes on the propagators, starting by the adiabatic modes. The effect of the isodensity mode $\Xi^{(\iis)}{}_a^{\ b}$ will be discussed in the next section.
The adiabatic modes in $\Xi_{a}^{\ b}$ can be simply reabsorbed in a phase transformation of the propagator  \cite{2012PhRvD..85f3509B}, or equivalently of the fields,
\[
\xi_{a}^{\ b}(\vk,\eta,\eta_0;\Xi^{({\rm ad})})=g_{a}^{\ b}(\eta,\eta_0)\exp\left(\int^{\eta}_{\eta_0}\dd\eta'\Xi^{({\rm ad})} (\vk,\eta')\right)\;. \label{resummed_1}
\]
This phase transformation affects the amplitude of the average propagator $\hxi_{a}^{\ b}(k,\eta,\eta_0) \equiv \langle \xi_{a}^{\ b}(\vk,\eta,\eta_0;\Xi^{({\rm ad})})\rangle_{\Xi} $ \footnote{The average propagator $\hxi_a^{\ b}(k, \eta,\eta_0)$ coincides with what is usually defined as the nonlinear propagator $G_{ab}(k, \eta,\eta_0)$ in \cite{2008PhRvD..78j3521B,2010PhRvD..82h3507B,2012PhRvD..85f3509B}.}, which reads
\[
\hxi_{a}^{\ b}(k,\eta,\eta_0)=g_{a}^{\ b}(\eta,\eta_0)\exp\left(\sum_{p=2}^{\infty}\frac{c_{p}}{p!}\right)\;, \label{G_cumulants}
\]
where $c_{p}$ is the $p$-order cumulant of the field $  \int^{\eta}_{\eta_0}\dd\eta'\Xi(\vk,\eta')$. 
This is at the origin of the exponential cutoff originally found by Crocce and Scoccimarro in \cite{Crocce:2005xy}. The form (\ref{resummed_1}) is however 
specific of adiabatic large-scale modes, as we stressed in \cite{2012PhRvD..85f3509B}.

\subsection{Equal time multi-spectra}

In the previous subsection we specifically focused our interest on the properties of the propagators. 
Let us now turn our attention to equal time
spectra. We first note that,  since in (\ref{spectraAdiab}) $\xi_a^{\ b}(-\vk,\eta,\eta_{0};\Xi^{\rm ad}_{-\vk})=\xi^{*}\!{}_a^{\ b}(\vk,\eta,\eta_{0};\Xi^{\rm ad}_{\vk})$, \textsl{
the linear power spectra of the eikonal system are independent on the large-scale adiabatic  modes}.
Thus, in the absence of isodensity modes, the theory constructed with the eikonal approximation shares the same linear power spectra as standard PT, even though their propagators are different.
In this paragraph we want to explicitly show that this property extends to multi-spectra and is valid not only  at leading
order, but also  at any given specific order in standard PT.

To show this, let us first  consider the 1-loop result  computed in the previous section, in particular Eq.~(\ref{onelooplk}).
This equation explicitly  gives the leading large-$k$  behavior of the 1-loop correction, which is proportional to $k^2$, and shows that adiabatic modes do not contribute to the 1-loop correction of the power spectra  at this order in $k$. We have also shown that this  result can be understood at a diagrammatic level as a cancellation between the adiabatic contributions to the loops in the soft limit.

The arguments of that section automatically extend to the eikonal approximation. Let us consider the three contributions to the 1-loop correction to the power spectra depicted on Fig.~\ref{AdiabEikos}, where now we assume that the dashed lines that appear in the diagrams are soft modes in the eikonal limit
and that they carry only adiabatic modes, whose specific time dependence 
is irrelevant in this discussion. By Eq.~\eqref{Xi_3}, the contribution of these modes to each diagram 
is the insertion of a  diagonal leg,
\begin{equation}
\Xi_{a}^{\ b}(\vk,\eta)=\Xi^{({\rm ad})} (\vk,\eta)  \; \delta_{a}^{\ b} \;,
\label{eikovertex}
\end{equation}
on the principal line. Because this insertion is diagonal, the cancellation of the first two diagrams with the last one follows, similarly to what was discussed in the previous section, see Eq.~\eqref{eqc} in Sec.~\ref{sec:c}.

\begin{figure}
\centerline{\epsfig {figure= 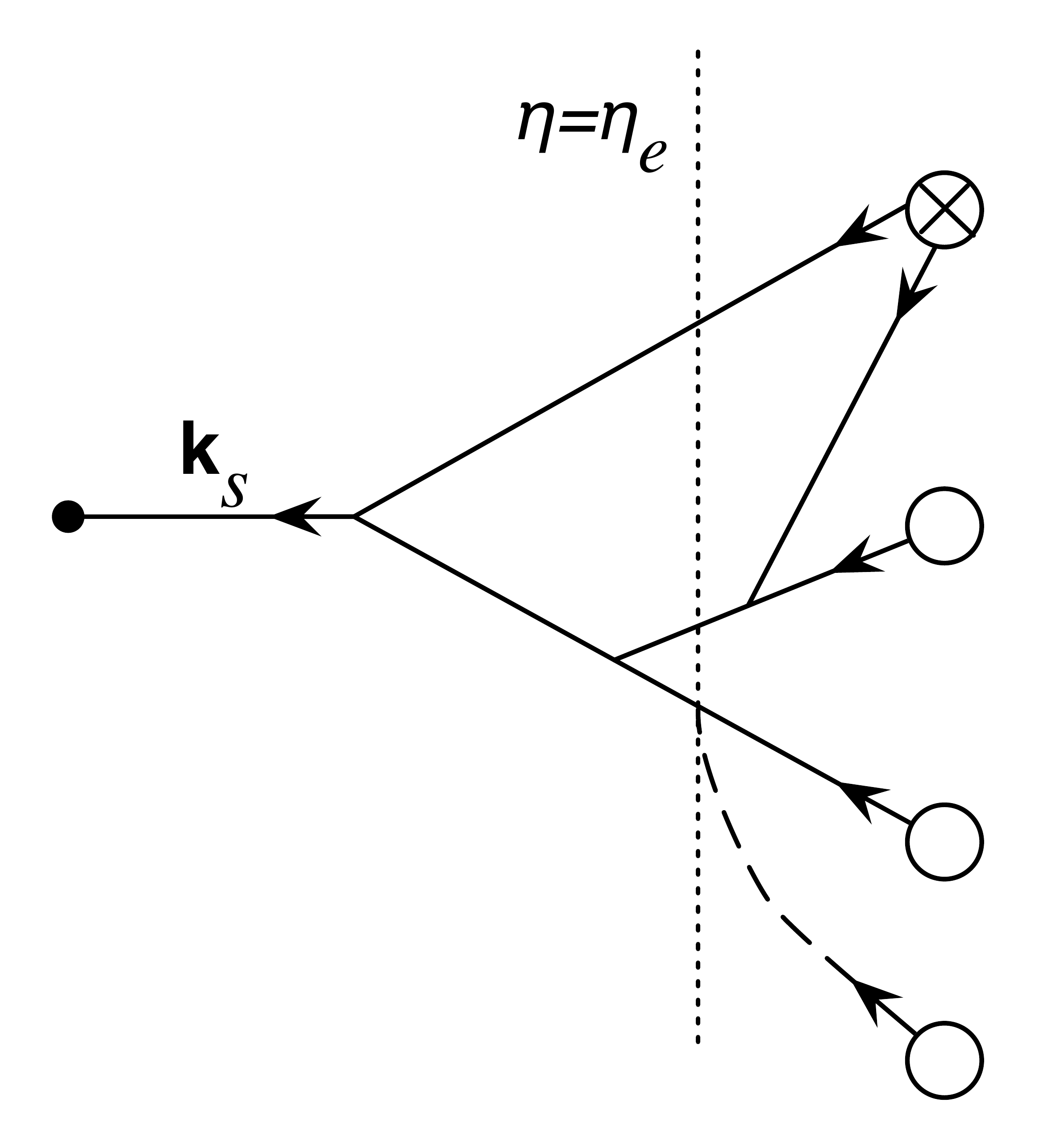,width=4.5cm}}
\caption{The solid lines in this diagram represent one contribution to the 1-loop correction to the 3-point propagator. The diagram has been
drawn in such way that vertical (dotted) lines cross propagator lines at the same time. The dashed line represents the impact of an incoming soft
mode.}
\label{EikonalTheorem2}
\end{figure}

We can now extend these results to any diagram. As an example, consider the diagram of Fig.~\ref{EikonalTheorem2} which, in absence of the insertion of the dashed line, represents a 1-loop contribution to the 3-point propagator. 
It has been drawn in such a way that vertical lines cross the diagram at  the same time. For instance, the dotted line denotes $\eta=\eta_e$. First, we want to study the impact of coupling this diagram to \textsl{adiabatic} modes in the eikonal limit. To do that, we need to sum over all possible insertions of $\Xi^{\rm ad}(\vk_{i},\eta_{e})\delta_{a}^{\ b}$ to any of the lines of the diagram with wavemode $\vk_i$ and integrate over $\eta_e$.
We will do it step by step. 
\begin{itemize}
\item Consider a line of the diagram with wave number $\vk_{i}$ crossing $\eta_e$. We can insert the dashed line at $\eta_e$. Before integrating over $\eta_e$ the new diagram---the one obtained after insertion---is just proportional to the original one, because 
the insertion of the identity does not change the matrix structure of the diagram. 
The diagram is simply multiplied by $\Xi^{\rm ad}(\vk_{i},\eta_{e})$.
\item Then  one sums over all insertions at the same time $\eta_e$. Because the time is the same,
the time dependence factorizes out and one gets a proportionality factor given by
$\sum_{i_{e}} \Xi^{\rm ad}(\vk_{i_{e}},\eta_{e})$. Note that the wave modes that enter this sum
depend on $\eta_{e}$.
\item The sum of the wave modes, $ \sum_{i_{e}}\vk_{i_{e}}$, is independent of $\eta_{e}$: it is given
by $ \sum_{i_{e}}\vk_{i_{e}}=\vk_{s}$, i.e.~the total wave mode. As $\Xi^{\rm ad}(\vk_{i},\eta_{e})$ is proportional to $\vk_i$, the resulting proportionality factor can then
be  rewritten as $\Xi^{\rm ad}(\vk_{s},\eta_{e})$, which should be finally integrated over time.
\end{itemize}
As a result, the impact of a single interaction with modes in the eikonal limit is a multiplication by 
$\int_{0}^{\eta} \dd \eta_{e} \Xi^{\rm ad}(\vk_{s},\eta_{e})$. This result is valid for multipoint propagators computed at any loop order.

The computation of equal time (multipoint) spectra is obtained from the product of multipoint propagators in such a way that 
$\sum_{s}\vk_{s}=0$, to ensure statistical homogeneity. As a result, when one takes into account the impact of adiabatic large-scale modes 
on a diagram contributing to a multispectrum, one gets the corrective factor 
$\sum_{s}\int_{0}^{\eta} \dd \eta_{e} \Xi^{\rm ad}(\vk_{s},\eta_{e})$, which identically vanishes. 

What are the consequences of this cancelation? Consider  a diagram that contributes to an equal time (multipoint) spectrum. 
Then any 1-loop correction \footnote{Although we present the calculations for Gaussian
initial conditions,
the results derived here are valid irrespectively of the nature of the initial conditions, as our arguments do not depend on where the incoming line emerge from.} to this diagram vanishes in the eikonal limit for adiabatic modes: each end line of the loop that one wants to insert should indeed run across the diagram and the previous result directly applies. Moreover, by subsequent use of this property, any equal-time spectrum computed at a given order in standard PT, is independent of the large-scale adiabatic modes. In particular, it is  independent of $\sigma_{d} \equiv \sigma_{++}$. 
A direct consequence of this property
is that, although $\sigma_{d}$ might appear in some intermediate calculations, it should vanish in the final result, were we able to properly resum all
contributing terms \footnote{In \cite{2012arXiv1205.2235A}, a scaling relation has been put forward where the
nonlinear power spectrum depends only on the reduced wave-mode  $y \equiv e^{\eta} \sigma_d k$. While 
this definition approximately captures the time dependence of the power spectrum, its $\sigma_d$ 
dependence is in contradiction with our findings.}. Note that this  does not apply to any diagram in standard PT. It applies only to any
set of diagrams that contributes (at a given order in standard PT) to equal time multipoint spectra.

This property has been used in
\cite{Bernardeauetal1012a} to build regularized diagram values in such a way that they are explicitly independent of $\sigma_{d}$.
Note also that the result obtained here for standard PT is not necessarily verified for other perturbation schemes, where the number of the vertices of the contributing  diagrams is not constant at each order. For instance, this is the case of RPT, where $\sigma_{d}$ explicitly appears in the expression of the final power spectra. 

We finally stress that this property only holds for adiabatic modes; it is not respected by \textsl{isodensity} modes \cite{2012PhRvD..85f3509B}. 
Indeed,  as discussed in the next section,  isodensity modes not only affect the phase of the propagator but also its amplitude; 
consequently, they affect the amplitude of power spectra.


\section{The small-scale power spectrum for CDM and baryons in the eikonal approximation}
\label{sec:SectionIV}

In this section we compute the  power spectra for CDM and baryons on small scales and high redshift, when both adiabatic and nonadiabatic modes are 
taken into account. We start with the description of the mode evolution, explore the consequences of the presence of a finite sound speed
and then describe the resulting properties of the small-scale power spectra.

\subsection{Mode evolution}

\begin{figure}
\centerline{\epsfig {figure=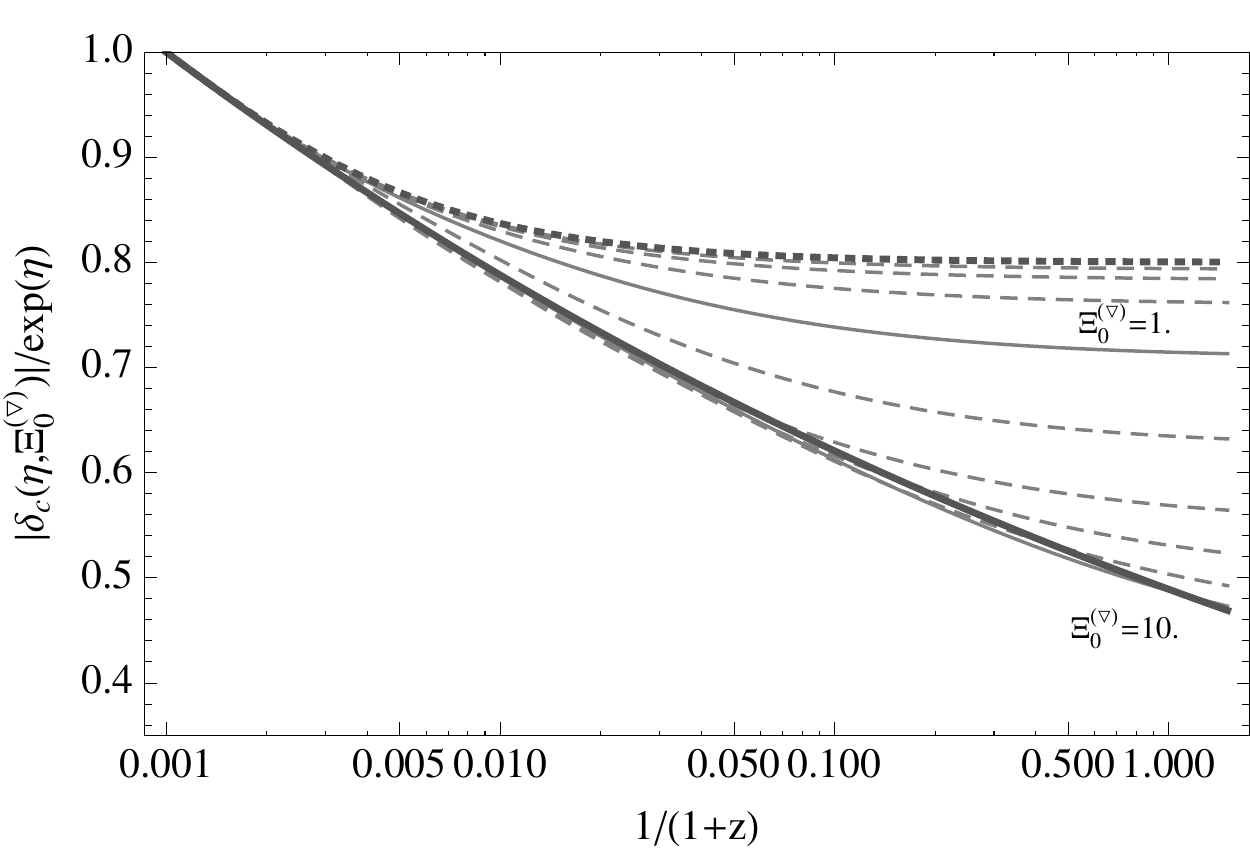,width=\columnwidth}}
\caption{The amplitude of the CDM density (in units of $\delta_{0}\exp(\eta)$) as a function of redshift $z$, for different $|\Xi_{0}^{(\iis)}|$, and assuming the initial conditions to be given at $z=1000$ by the growing mode of a pure CDM fluid with no baryon density or velocity fluctuations. The values of $|\Xi_{0}^{(\iis)}|$ are exponentially distributed, from $0.25$ (top) to $10.$ (bottom). The top thick dotted line
corresponds to the form of Eq.~(\ref{eq:asympc1}) and the bottom thick solid line to the form of Eq. (\ref{eq:asympc}).}
\label{EikonalDampingCDM}
\end{figure}

\begin{figure}
\centerline{\epsfig {figure= 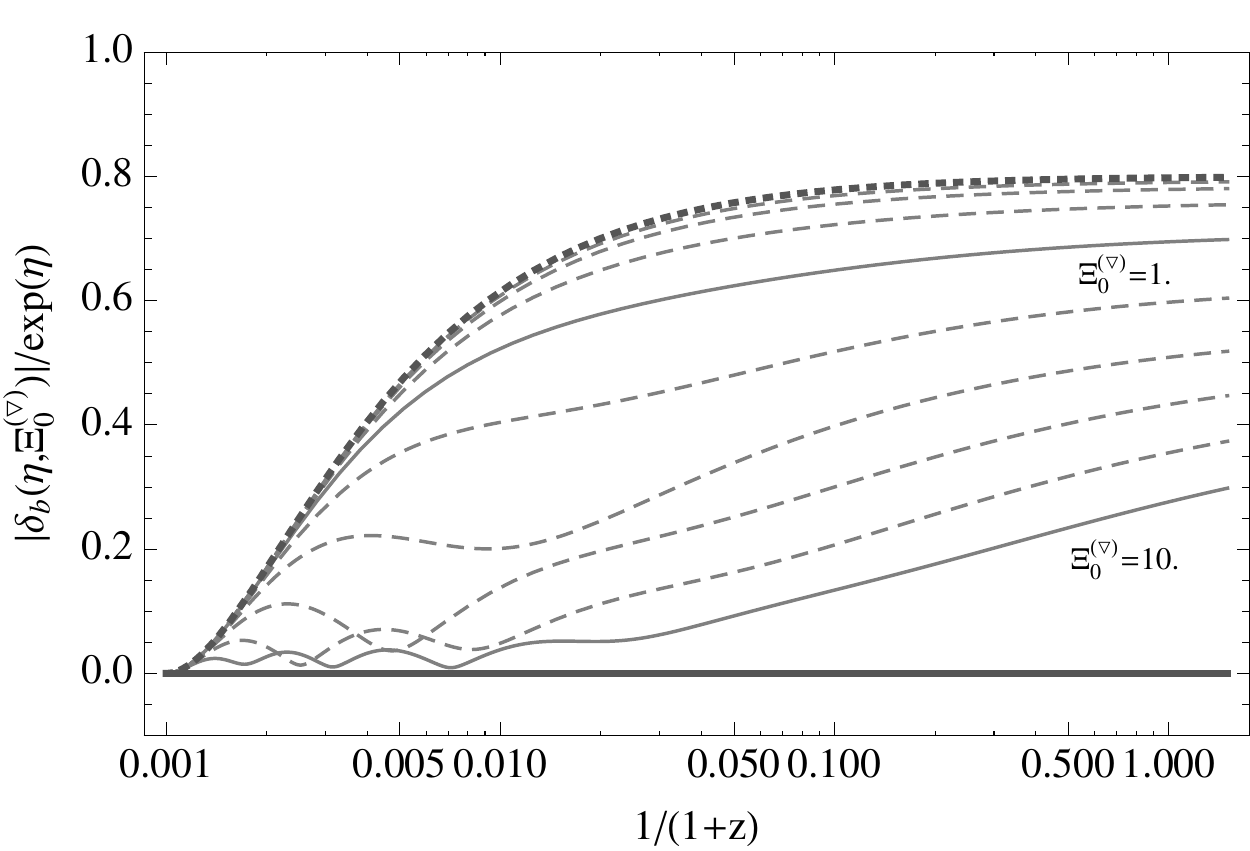,width=\columnwidth}}
\caption{The amplitude of the baryon density. Same conventions as in Fig.~\ref{EikonalDampingCDM} with appropriate theoretical predictions.}
\label{EikonalDampingBaryons}
\end{figure}

\begin{figure}
\centerline{\epsfig {figure= 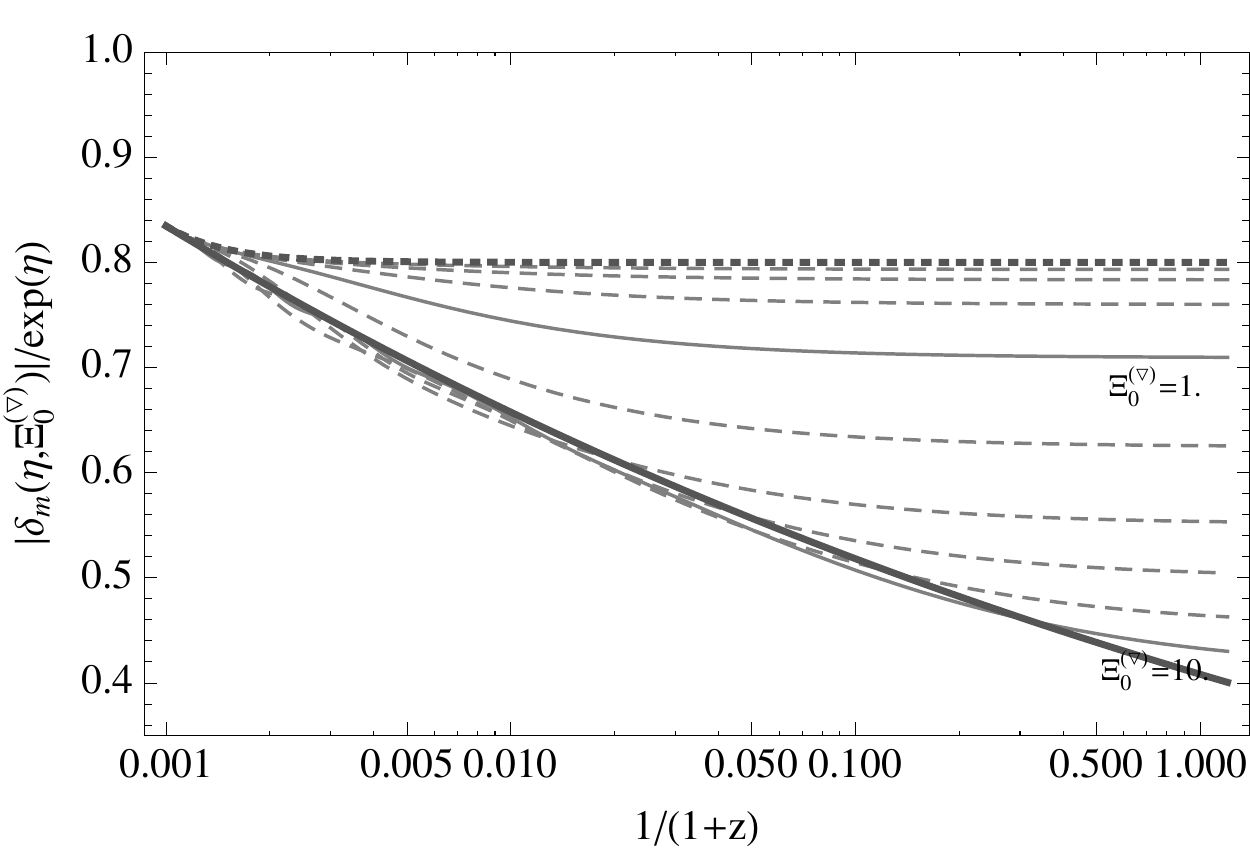,width=\columnwidth}}
\caption{The amplitude of the total matter density as given by Eq.~(\ref{eq:totalmassgrowth}). Same conventions as in previous Figs.~\ref{EikonalDampingCDM} and \ref{EikonalDampingBaryons}}
\label{EikonalDamping}
\end{figure}

Here we work out the linear solutions of the baryons-CDM fluid system in the eikonal approximation, when there
is a nonzero isodensity velocity field. It is simpler to rewrite \eqref{eikonalNL} in components; for the sake of clarity we also assume pure matter dominance, $\Omega_{\rm m}=1$ and $f_+=1$. Thus, dropping the right-hand side, for each mode $\vk$  this equation yields
\begin{align}
\frac{\partial }{\partial \eta} \delta_{a}(\eta)- \Theta_{a}(\eta)+\ii\alpha_{a}(\eta)\delta_{a}(\eta)&=0 \;, \label{cont_eik}
\\
\frac{\partial }{\partial \eta} \Theta_{a}(\eta) +\frac{1}{2}\Theta_{a}(\eta)+ \ii\alpha_{a}(\eta)\Theta_{a}(\eta)-
\frac{3}{2}\sum_{b}\fw_{b}\delta_{b}(\eta)&=0 \;,\label{eul_eik}
\end{align}
for $a= \cdm, \ba$, meaning CDM and baryons.
We have seen above that the long-wavelength adiabatic modes can be reabsorbed by a common phase redefinition. Thus, for simplicity
here we will assume that the large-scale flow is only in the isodensity modes. In this case, the  functions $\alpha_a$ in Eqs.~\eqref{cont_eik} and \eqref{eul_eik} are simply given by 
\be
\alpha_a \equiv  \ii \int \dd^3 \vq \frac{\vk \cdot \vq }{q^2} \; \Theta_a^{(\iis)} \;.
\label{alpha_te}
\ee
By Eq.~\eqref{XiisDef} these are related to $\Xi^{(\iis)}$ by
\be
\alpha_\cdm = \ii \fw_\ba \Xi^{(\iis)} \;,  \quad  \alpha_\ba = - \ii \fw_\cdm \Xi^{(\iis)}\;. \label{alpha_Xi}
\ee
Since $\Xi^{(\iis)}$ is  imaginary, these functions are real.

As discussed in  \cite{2012PhRvD..85f3509B}, we can gain insight into this system of equations by making the following change of variables,
\begin{align}
\hdelta_{a}(\eta)&= \delta_{a}(\eta) e^{\ii\ialpha_{a}(\eta)} \;, \\
\hTheta_{a}(\eta)&=\Theta_{a}(\eta) e^{\ii\ialpha_{a}(\eta)} \;,
\end{align}
where $\ialpha_{a}$ is the time integral of $\alpha_{a}$,
\[
\ialpha_{a}(\eta)=\int_{\eta_{0}}^{\eta} \dd \eta' \alpha_{a}(\eta')=2\alpha_{a}(\eta_0) \left(1 - e^{-(\eta-\eta_0)/2} \right) \;, \label{hatalpha}
\]
where for the second equality we have used Eq.~\eqref{alpha_te} and the time evolution of the isodensity mode,  $\alpha_a \propto \Theta_a^{(\iis)} \propto e^{-\eta/2}$.
Since $\alpha_a$ is purely real, this change
of variable is only a time-dependent  phase change. Note however that, contrary to the adiabatic case, this phase change is different for different species.

With these new variables, the system of equations \eqref{cont_eik} and \eqref{eul_eik} reads
\begin{align}
\frac{\partial }{\partial \eta} \hdelta_{a}(\eta)- \hTheta_{a}(\eta)&= 0\label{hatCont} \;, \\
\frac{\partial }{\partial \eta} \hTheta_{a}(\eta)+\frac{1}{2}\hTheta_{a}(\eta)- \frac{3}{2}\sum_{b}\fw_{b}\hdelta_{b}(\eta)e^{\ii(\ialpha_{a}-\ialpha_{b})}&= 0 \;.
\label{hatEuler}
\end{align}
The exponential in the third term of Eq.~\eqref{hatEuler} affects the coupling between CDM and baryons. When the indices $a$ and $b$  in the exponent are the same, the phase difference vanishes and there is no effect; for different indices the phase difference is proportional to
\[
\ialpha_{\cdm}(\eta)-\ialpha_{\ba}(\eta)   = 2  \ii 
\; \Xi^{(\iis)}_0 \left(1 - e^{-(\eta-\eta_0)/2} \right)  \;, \label{phase_diff}
\]
where $ \Xi^{(\iis)}_0 \equiv \Xi^{(\iis)} (\eta_0)$ and the equality follows from Eqs.~\eqref{hatalpha} and \eqref{alpha_Xi}.
Thus, the solution of the system (\ref{hatCont}) and (\ref{hatEuler}) can be described as a function of time and $ \Xi^{(\iis)}_0 $, which parametrizes the phase difference due to the relative velocity between CDM and baryons. As we will see, if the phase difference changes rapidly in time with a rate of change faster than that of structure formation, the coupling between the two species can be  neglected, i.e.~the two species effectively decouple.

Let us now concretely solve this system of equations. For simplicity we  assume  that initially the baryons do not exhibit significant fluctuations at the scale of interest,  $\hat \delta_\ba (\eta_0)=0$. Equation~\eqref{cont_eik} implies that also their velocity vanishes.
The physical question that we want to address is then to describe how the baryons fall into the CDM potential wells.
For the CDM we assume that it is initially in
the single fluid growing mode. As the baryons are unperturbed, this is given by \cite{1980PhRvL..45.1980B,2012PhRvD..85f3509B}
\be
\begin{split}
\hdelta_{\cdm}(\eta)&\propto \exp(-\nu_{\cdm}\eta) \;, \\
\hTheta_{\cdm}(\eta)&=\nu_{\cdm}\hdelta_{\cdm}(\eta) \;,
\end{split}
\ee
with
\[
\nu_{\cdm}=\frac{\sqrt{1+24 \fw_{\cdm}}-1}{4}\;.
\]

In the absence of large-scale isodensity perturbations, $\Xi^{(\iis)}_0 =0$ and $\hdelta_a = \delta_a$, $\hTheta_a = \Theta_a$. In this case the two fluids evolve  according to the linear theory and the solutions of Eqs.~\eqref{hatCont} and \eqref{hatEuler} are then
\begin{widetext}
\begin{align}
{\delta_{\cdm}(\eta, \Xi^{(\iis)}_{0}=0)}/{ \delta_{\cdm}(\eta_0)} =& \
\frac{\fw_\cdm (3 + 2 \nu_\cdm)}{5} e^{\eta -\eta_0 } -\frac{ 2 \fw_\cdm (\nu_\cdm -1)}{5} e^{-3 (\eta -\eta_0) /2}
-2 { \fw_\ba \nu_\cdm} e^{-(\eta -\eta_0) /2}+\fw_\ba (1+2 \nu_\cdm)
  \;,\label{eq:asympc1}
   \\
{\delta_{\ba}(\eta, \Xi^{(\iis)}_{0}=0)}/{\delta_{\cdm}(\eta_0)}
=&\ \frac{\fw_\cdm (3+2 \nu_\cdm)}{5} e^{\eta -\eta_0 }-\frac{2 \fw_\cdm (\nu_\cdm -1)}{5}   e^{-3 (\eta -\eta_0) /2}
+2 \fw_\cdm \nu_\cdm e^{-(\eta -\eta_0) /2} - \fw_\cdm (1+2 \nu_\cdm)
      \;.\label{eq:asympb1}
      \end{align}
\end{widetext}
On the other hand, for asymptotically large values of $\Xi^{(\iis)}_{0}$ the relative phase oscillation between the two fluids is  large and they do not ``see'' each other and evolve independently.
In this case
\begin{align}
{\hdelta_{\cdm}(\eta, \Xi^{(\iis)}_{0}\to \infty)}/{ \hdelta_{\cdm}(\eta_0)}& = \exp(-\nu_{\cdm}(\eta-\eta_{0})) \;, 
\label{eq:asympc}\\
{\hdelta_{\ba}(\eta, \Xi^{(\iis)}_{0}\to \infty)}/{ \hdelta_{\cdm}(\eta_0)}& \to 0 \;.
\label{eq:asympb}
\end{align}

Finally, for finite values of $\Xi^{(\iis)}_{0}$ we have to rely on numerical integration.
Note that in general the resulting matter transfer function is given by
\[
\begin{split}
&\vert\delta_{\rm m}(\eta,\Xi^{(\iis)}_0)\vert
\\
&\ =
\vert \fw_{\cdm}\hdelta_{\cdm}(\eta,\Xi^{(\iis)}_0)
+ \fw_{\ba} e^{\ii (\hat \alpha_\cdm(\eta) - \hat \alpha_\ba (\eta))}
\hdelta_{\ba}(\eta,\Xi^{(\iis)}_0)\vert
\\
&\ =
\vert \fw_{\cdm}\delta_{\cdm}(\eta,\Xi^{(\iis)}_0)
+\fw_{\ba}\delta_{\ba}(\eta,\Xi^{(\iis)}_0)\vert\;.
\label{eq:totalmassgrowth}
\end{split}
\]
The amplitude of the CDM, baryons and total matter density transfer functions are shown on Figs.~\ref{EikonalDampingCDM}--\ref{EikonalDamping}. We clearly see the damping due to the isodensity modes and its evolution with the redshift for various values of $\Xi^{(\iis)}_{0}$. 
When $|\Xi^{(\iis)}_{0}| \la 1$ the impact of the isodensity modes is limited---the evolution of the density contrasts approaches that of Eqs.~(\ref{eq:asympc1}) and (\ref{eq:asympb1}) shown as thick dotted lines---whereas
the asymptotic forms (\ref{eq:asympc}) and (\ref{eq:asympb}), represented as thick solid lines on Figs.~\ref{EikonalDampingCDM}--\ref{EikonalDamping},  are clearly followed for  $|\Xi^{(\iis)}_{0} |\ga 10$ until the present time. This represents the maximum damping in the total mass density contrast that the absence of growth in the baryon component can generate~\footnote{It can be noted that the total density contrast in Fig.~\ref{EikonalDamping}  can be  occasionally below this asymptotic curve. This is due to the fact that the CDM and baryon density contrasts are transitorily in opposition of phase.}.

\subsection{Effects of a finite sound speed}
\begin{figure}
\centerline{\epsfig {figure=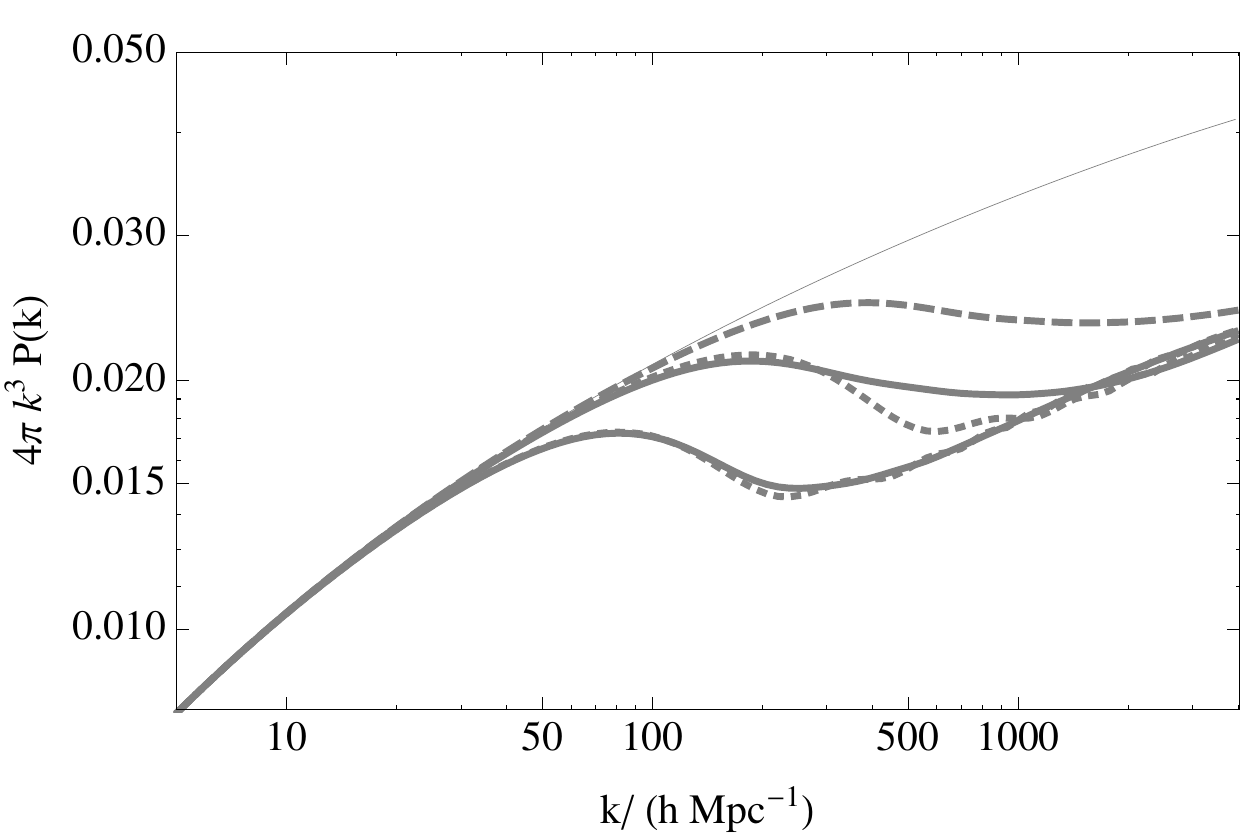,width=8cm}}
\caption{Impact of a finite sound speed and of long-wavelength isodensity modes
on the total matter power spectrum, computed at $z=40$. The thin solid line represents the linear power spectrum
in absence of sound speed and mode damping effects. The spectrum including a finite sound speed given by Eq.~(\ref{csexpress}) is shown by the dashed line. The two dotted lines represent the power spectrum with zero sound speed but including the effect of mode damping due to relative velocity for two different values of the angle between the wave mode $\vk$ and the relative displacement field, i.e.~$u= 0.5$ and $u= 0.2$  (see Eq.~\eqref{u}), respectively from top to bottom.  
Finally, the two solid lines include both effects of mode damping and finite sound speed. The amplitude of the mode damping effect corresponds
to what is expected in the Local Group environment, as described in subsection \ref{anispecandcv}.}
\label{DampCsEffects}
\end{figure}

In practice, the scales where the damping effect takes place is not far from the scales where the finite sound speed of baryons starts playing a significant role.
In this case one needs to change the Euler equations for the baryons into
\[
\Theta_{\ba}'(\eta)+\frac{1}{2}\Theta_\ba(\eta)- \frac{3}{2}\sum_{a}\fw_{a}\delta_{a}(\eta)+ \frac{k^{2} c_{s}^{2}(\eta)}{\HH^2 (\eta)}\delta_\ba(\eta)=0 \;,
\]
where $c_{s}(\eta)$ is the time dependent sound speed. For its value we adopt  the expression used in \cite{2010PhRvD..82h3520T},
\[
c_{s}^{2}=\frac{\gamma k_{\rm B} T_\ba}{\mu m_{\rm H}} \;,
\label{csexpress}
\]
where $\gamma=5/3$ for an ideal monatomic gas, $k_{\rm B}$ is the Boltzmann constant, $\mu=1.22$ is the mean molecular weight including a helium mass fraction of $0.24$, $m_{\rm H}$ is the mass of the hydrogen atom and $T_\ba$ is the kinetic temperature of the baryons.
The latter is determined by a competition between adiabatic cooling and Compton heating from the CMB. Following \cite{2010PhRvD..82h3520T}, we
parametrize it as
\[
T_\ba (a)=\frac{T_{\rm CMB,0}}{a}\left[1+\frac{a/a_{1}}{1+(a_{2}/a)^{3/2}}\right]^{-1} \;,
\]
with $a_{1}=1/119$, $a_{2}=1/115,$ and $T_{\rm CMB,0}=2.726$ K.

The impact on the total matter power spectrum of a finite sound speed and of long-wavelength isodensity modes between CDM and baryons is presented in Fig.~\ref{DampCsEffects}. 
The thin solid line represents the linear matter power spectrum in absence of both effects. The dashed line shows the effect of a finite sound speed only, whereas the dotted lines show the effect of the isodensity modes
with two different values of the angle between the wave mode $\vk$ and the relative displacement field. Finally, the two solid lines show the power spectrum including both effects.

For realistic velocity differences, as discussed in more details in the next subsection, the relative velocity dominantly contributes to the damping effect for modes 
between $100\,h/\Mpc$  and $500\,h/\Mpc$.
The two 
predictions join when the structure growth in the baryon growth is fully suppressed irrespective of the presence
of the isodensity mode. This happens for wave modes above  
roughly $1000\,h/\Mpc$. The resulting spectrum when the two effects are taken into account together (solid lines) is therefore very close
to the one obtained from relative velocity damping only.
For realistic velocity differences, as defined in the next subsection, the relative velocity effect is the dominant  damping effect for modes 
between $100\,h/\Mpc$  and $500\,h/\Mpc$. 

It is important to stress that these two effects, although they are closely related, have very different phenomenological signatures:
the damping due to a finite baryon sound speed is  uniform and isotropic; the damping due to the relative velocity between CDM and baryons is spatially modulated and anisotropic. We explore in more details the consequences of these latter properties in the next paragraph.

\subsection{Anisotropic spectrum and cosmic variance}
\label{anispecandcv}

\begin{figure}
\centerline{\epsfig {figure=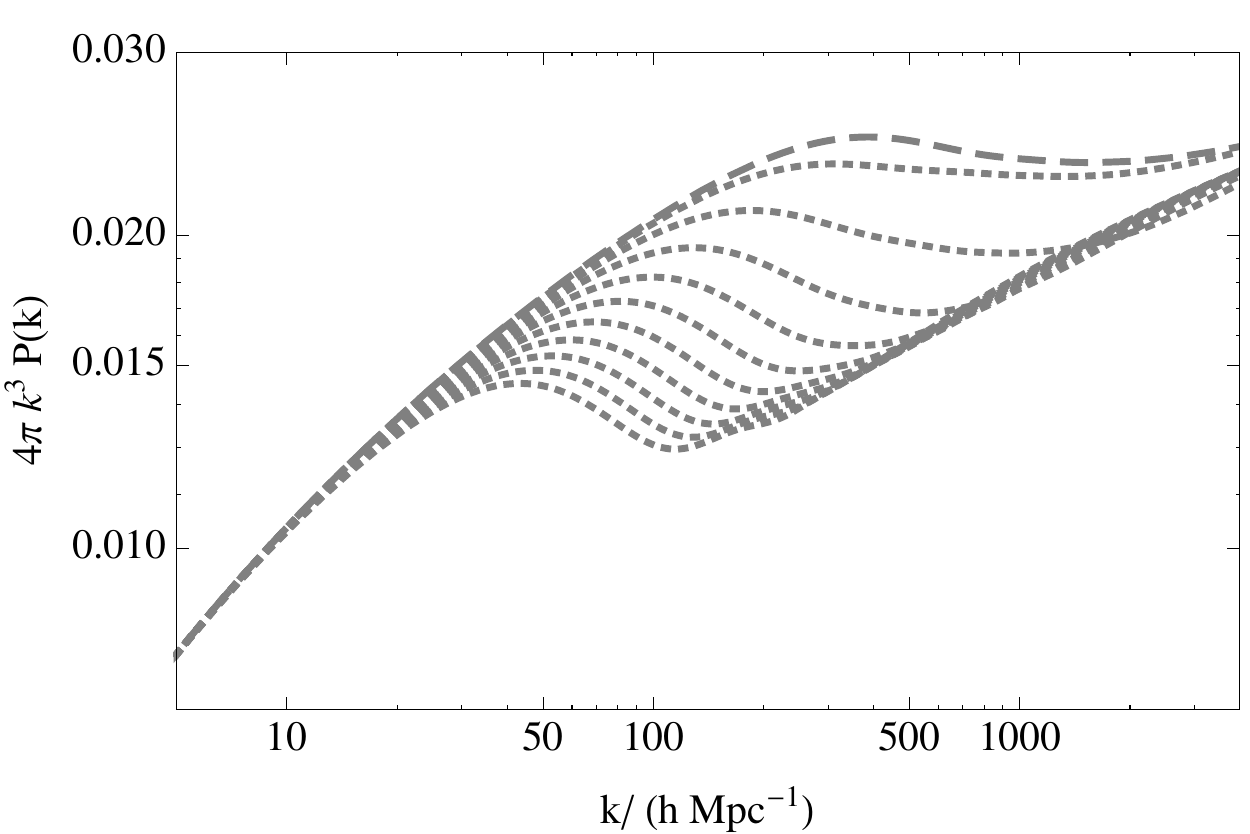,width=8cm}}
\caption{Matter power spectrum at $z=40$ in the Local Group environment  as a function of $k$, for different values of $u_{k} \equiv  \hat \vk \cdot \hvdi$ varying from $0.1$ to $1$ (dotted lines from top to bottom). The
dashed line corresponds to absence of effects, i.e.~$\vk$ orthogonal to $\vd^\iis$. The amplitude of the relative displacement corresponds to what is expected in the Local Group, $d^{\iis}_{\rm LG}=30.5\ h^{-1}\Kpc$, see Eq.~\eqref{valuedLG}.}
\label{PowerAnis40}
\end{figure}

\begin{figure}
\centerline{\epsfig {figure=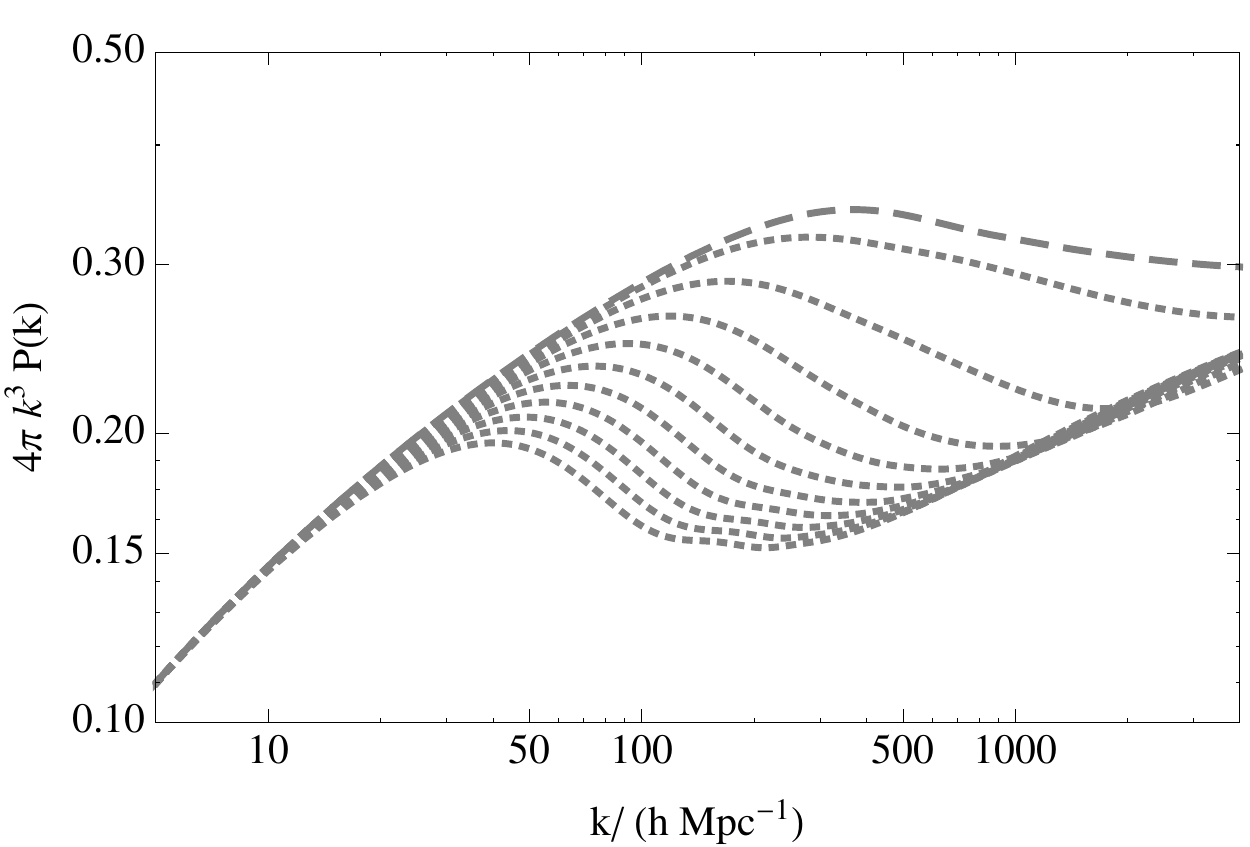,width=8cm}}
\caption{Same as  Fig.~\ref{PowerAnis40} for z=10.}
\label{PowerAnis10}
\end{figure}

We define the (comoving) relative displacement between the CDM and baryons, integrated in time, as
\be
\vdi \equiv  \int_{\eta_{0}}^{\eta}\dd\eta'\int\dd^{3}\vq\frac{\ii \vq}{q^{2}}\Psi_{\iis} (\vq,\eta)  \;,
 \label{defdlg}
\ee
where we can take the upper bound of the time integral as large as we want, in such way that the integral becomes time independent.

For a fixed value of $\vdi$ the power spectrum of the density field  is anisotropic. It explicitly depends on
the value of the parameter $\Xi^{(\iis)}$ defined in \eqref{XiisDef}, related to $\vdi$ by
\[
\Xi^{(\iis)}=- {\ii} \vk \cdot \vdi\equiv - {\ii} u \, k \, \ndi \;.
\label{u}
\]
To illustrate the dependence of the power spectrum on this parameter we compute  the expected amplitude of $\ndi$
in the Milky Way environment or, more generally, in the Local Group environment.

The Local Group peculiar velocity can be expressed in terms of the initial modes (at time $\eta_{0}$). In particular, using that the dimensionless velocity field $\Theta$ is related to the peculiar velocity by $\Theta = -\nabla \cdot \vv/(a H f_+)$, we can express the Local Group velocity as a function of the Fourier modes of the long-wavelength adiabatic modes,
\be
\vv_{\rm LG}(\eta)= f_{\plus}(\eta)H(\eta) a (\eta)e^{\eta-\eta_{0}}\int\dd^{3}\vq\frac{{\ii \vq}}{q^{2}}\Psi_{\plus}(\vq,\eta_{0}) \;.\label{vLG}
\ee
The same can be done for  the isodensity integrated displacement field,
\be
\vd^{\iis}_{\rm LG}(\eta)= - 2  \left(e^{-(\eta-\eta_{0})/2}-1 \right)\int\dd^{3}\vq\frac{{\ii \vq}}{q^{2}}\Psi_{\iis}(\vq,\eta_{0})\;,
\label{dLG}
\ee
obtained from Eq.~\eqref{defdlg} by time integration of the peculiar velocity field due to the isodensity modes, whose time dependence is $\Psi_\iis \propto e^{-\eta/2}$.  As a result, $d_{\rm LG}(\eta)$
is not entirely determined by the observed velocity field of the Local Group but it is however expected to be strongly related to it.

We can compute the expected value of the displacement field today, given the observed value of the Local Group velocity flow. Assuming that $\Psi_+$ and $\Psi_\iis$ are Gaussianly distributed, one finds that the expectation value of $\Psi_\iis$ given $\Psi_+$ is $\langle \Psi_\iis \rangle_{\Psi_+}= (\sigma^2_{+ \iis} / \sigma^2_{++}) \Psi_+$.
Using Eqs.~\eqref{vLG} and \eqref{dLG} at late time,  the displacement field (in physical unit) today  is given by
\[
\mg \vd^{\iis}_{\rm LG}\md_{v_{\rm LG}}=\frac{2}{f_{\plus}} \frac{D_+ (z_0)}{D_+(0)} \frac{\sigma^{2}_{\plus\iis}}{\sigma^{2}_{\plus\plus}} \frac{\vv_{\rm LG}}{100 \,  {\rm km/s}}\, h^{-1} \Mpc \;, 
\]
where $\vv_{\rm LG}$ and $f_+$ are evaluated at $z=0$ and  the variables $\sigma_{\alpha\beta}$ are defined in Eq.~(\ref{sigmaalphabetadef}).
Note that ${\sigma_{\plus\iis}}/{\sigma_{\plus\plus}}\approx 0.95$, which indicates that the peculiar velocity we experience is essentially due to
a dark matter potential well. As a consequence, the cosmic variance of $\vd^{\iis}_{\rm LG}$ given the observed
velocity should be small. For each component $i$ it is given by
\[
\Delta (d^{\iis}_{\rm LG})^{i} = 2 \sigma_{\iis\iis} \, \sqrt{1-r^{2}}\label{deltadis} \;,
\]
where $r \equiv \sigma_{+ \iis}^2/(\sigma_{++} \sigma_{\iis \iis})$ is the correlation coefficient between the two modes. 
Using an estimated peculiar velocity of our Local Group of $627\pm22$ km/s  from the
CMB dipole measurement \cite{1993ApJ...419....1K}, we finally obtain 
\[
\vd^{\iis}_{\rm LG} \cdot \hat{\vv}_{\rm LG}\approx 30.5\pm 1 \pm 6.5 \ h^{-1} \Kpc\;  \label{valuedLG} 
\]
along the measured velocity. 
The first error comes from the uncertainty in the velocity determination and the second from cosmic variance 
(i.e. Eq.~(\ref{deltadis})), with identical cosmic variance error in the two transverse directions.

\begin{figure}
\centerline{\epsfig {figure=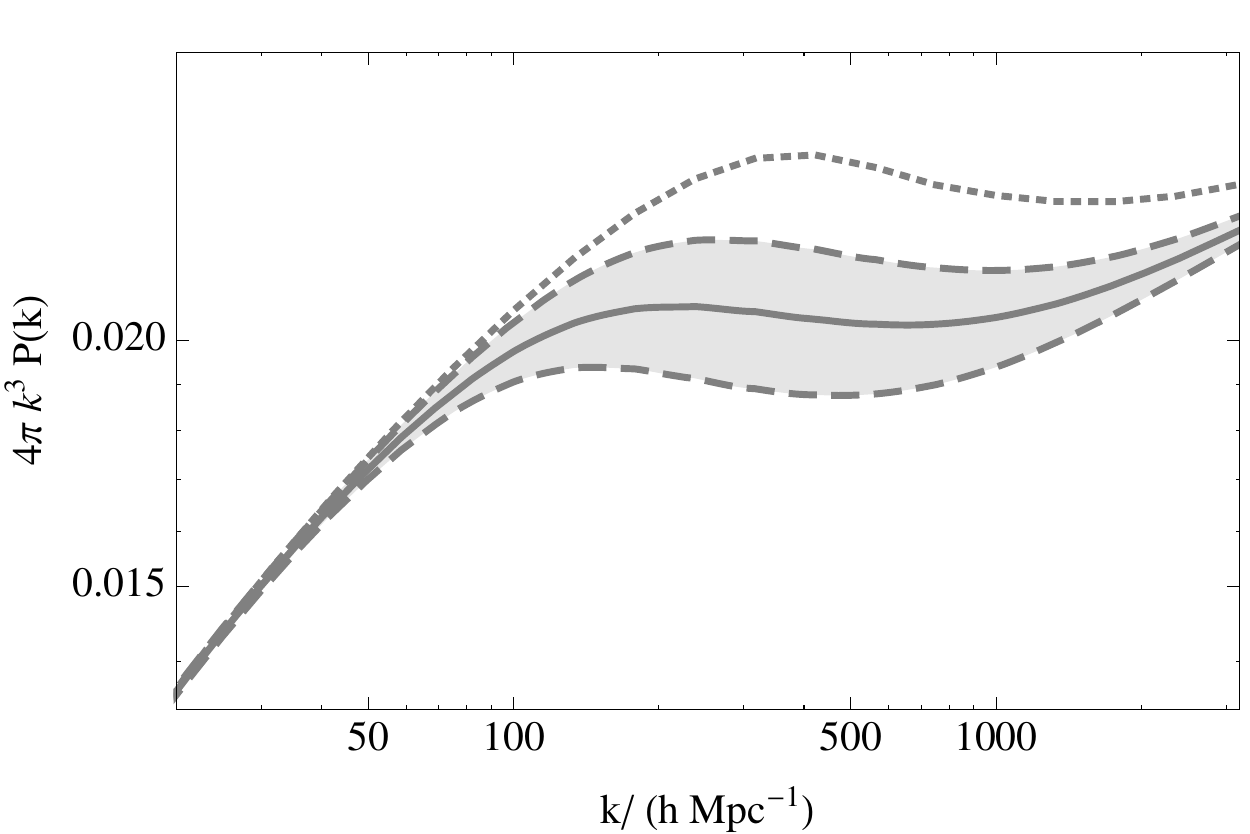,width=8cm}}
\caption{Expectation value and variance of the matter power spectrum at $z=40$.  The solid line is the expectation value, $\hP(k)$; the dashed lines show the expected 1$\sigma$ variation due to the fluctuations of the large-scale displacement field, i.e.~$\hP(k)\pm \DP(k)$, and the dotted line is the standard linear power spectrum. The calculations have been done for the standard concordant model.
}
\label{PowerCVar}
\end{figure}

We illustrate the implication of this result in Figs.~\ref{PowerAnis40}--\ref{PowerAnis10}, where we explicitly show the dependence of the matter power spectrum $P_{\delta}(\vk)$ on the relative angle $u$, while the amplitude of the displacement field has been chosen to fit the observed velocity of the Local Group. At scales of interest we observe very large variations of the amplitude of the power spectrum: effectively the modes along the displacement  field are damped, while the modes in a transverse direction are not.
This illustrates the importance of the anisotropic shape of the power spectrum due to this effect.

In general, in a local cosmological region one thus expects the power spectrum to differ significantly from the standard linear
solution. One can further define an effective \textsl{isotropically averaged} power spectrum  obtained from its angular average \cite{2010PhRvD..82h3520T},
\[
\eP(k,\ndi) \equiv \frac{1}{2}\int_{-1}^{1}\dd u P(k,\vdi)\;.
\label{Peff}
\]
The ensemble average power spectrum is then the average of $\eP(k,\ndi)$ over the distribution of $\ndi$, $\mP(\ndi)$, which again we assume here to be a Gaussian. Thus 
\[
\hP(k)=\int_{0}^{\infty}\dd\ndi\,\eP(k,\ndi)\,\mP(\ndi)\;;
\]
from Eq.~\eqref{Peff} we can also compute its variance,
\[
\DP(k)=\left[\int_{0}^{\infty}\dd\ndi\,\eP^{2}(k,\ndi)\,\mP(\ndi)-\hP^{2}(k)\right]^{1/2}\;,
\]
due to the fact that the environment of the small-scale modes is itself a random quantity.
This is illustrated on Fig.~\ref{PowerCVar} where we show the ensemble average matter power spectrum at $z=40$ (solid) and its variance (dashed). The solid line is to be compared with the result of Ref.~\cite{2010PhRvD..82h3520T}. 

We end this section by  stressing that 
the 1-loop result of Sec.~\ref{sec:MultiPT} gives only the damping scale, $k_{\rm damp}$, but not the overall shape of the power spectrum at large $k$. The fact that the damping saturates is a truly nonperturbative effect, fully captured by the eikonal approximation.

\section{Conclusions}

In this paper we have explored the impact of the presence of multiple fluids on power spectra beyond linear
order. In particular, we have shown that isodensity modes---more specifically modes involving large relative velocities---induce a coupling between  large-scale and small-scale modes. On the very small scales, these couplings can be dominant.

In the first part of the paper we have investigated these effects using standard PT. In particular, we have derived Eq.~(\ref{onelooplk}), which describes how isodensity modes affect the small-scale shape of the power spectrum.
In the second part, we have employed the eikonal formalism to compute the impact of the large-scale modes on power spectra in a nonperturbative way.
The first important result that we have derived is that
adiabatic large-scale velocity modes do not change the final amplitude of the small-scale density fluctuations. 
More specifically, we have shown that large-scale adiabatic modes do  not contribute to multipoint equal-time spectra 
computed at any given order in standard PT.

Rewriting the dynamical equations in the eikonal approximation allows to compute the impact of large-scale isodensity modes nonperturbatively. 
Indeed, \textsl{relative velocity} modes change the dynamics in a nontrivial way: the growth of structure is  anisotropic and the average final amplitude is modulated by the large-scale environment. We illustrate this mechanism by computing the resulting renormalized power spectrum in the Local Group environment. Note that the eikonal approximation used in this context put predictions on a very solid ground. In particular, the ambiguity concerning the separation of scales in the use of the eikonal approximation for the propagators, signaled in \cite{2012PhRvD..85f3509B},  disappears. Indeed, scales contributing to the large-scale modes are in the range 0.01--0.1 $h/\Mpc$ whereas the wave modes that we want to describe are in the 50--500 $h/\Mpc$ range. Thus, here  the separation of scales is rather clear.

In the last section of the paper we have thoroughly  investigated the impact of isodensity modes on the matter power spectrum.
We have found that a large-scale relative velocity induces a significant effect at very high $z$. The amplitude of this effect evolves 
with $z$, as the damping gets more pronounced for smaller redshifts. The scale at which 
it takes place is independent of the redshift and can be derived from the 1-loop calculation, see Eqs.~\eqref{kdampval2} and (\ref{kdampval}). 

One of the main consequences of such a mechanism is that large-scale galaxy fields are potentially biased in a non-trivial way. Indeed, it is  possible to have a large  modulation of the number density of galaxies which is related to the isodensity modes. Biasing is potentially nonlocal, i.e.~scale dependent up to bulk flow scales, which are of the order of 100 $\Mpc$. As far as galaxy biasing is concerned, this may be potentially worrisome. However, investigations have shown that such an effect is small and can be controlled at BAO scales \cite{2011JCAP...07..018Y}.

We also stress here that the resulting power spectrum is not only damped but also locally anisotropic. 
Thus, we expect that 
on small scales and for an environment such as  the Milky Way, the first structures form in an effectively 2D dynamics as modes along the displacement field are strongly damped. As a consequence, the impact of the isodensity modes cannot  be \textsl{a priori} captured by a mere change of the usual approaches, such as the Press-Schechter formalism,  which are all based on initial isotropic random fields. These issues are also discussed in  \cite{2010PhRvD..82h3520T}.
\\
\\
{\bf Acknowledgements:} We thank Mart\'in Crocce, Massimo Pietroni, Rom\'an Scoccimarro and Patrick Valageas for interesting discussions. This work is partially supported by the French  Programme National de Cosmologie et Galaxies. 

\newcommand{\treeGamma}{\,^{\rm tree}\Gamma}

\appendix
\section{Multipoint propagators}
Multipoint propagators, introduced in \cite{2008PhRvD..78j3521B}, are defined as the ensemble average of functional derivatives of the 
cosmic field components with respect to the initial conditions,
\begin{equation}
\begin{split}
{\Gamma}_{a}^{\ \,b_{1}\dots b_{p}}(\vk,\vk_{1},\dots,\vk_{p};\eta,\eta_{0})\,
\delta_{\rm Dirac}(\vk-\sum_{i}\vk_{i}) \equiv \\
\left\langle \frac{\partial^{p} \Psi_{a}(\eta)}{\partial \Psi_{b_{1}}(\eta_{0})\dots \partial \Psi_{b_{p}}(\eta_{0})}\right\rangle_{c} \;.
\end{split}
\end{equation}
By construction, they are symmetric under any exchange of the pairs $\{b_{i},\vk_{i} \} \leftrightarrow \{ b_{j},\vk_{j} \}$, for $i\ne j$.

The 2- and 3-point propagators are used, at tree order, to compute the 1-loop power spectra  in Sec.~\ref{sec:MultiPT} of this paper, see Eq.~\eqref{1loopcorr}.
\textsl{At tree order}, their explicit expressions are given by the \textsl{symmetrized} form of 
\begin{equation}
\begin{split}
{\Gamma}_{a}^{\ \,b_{1}b_{2}}(\vk,\vk_{1},\vk_{2};\eta,\eta_{0})=\int_{\eta_{0}}^{\eta}\dd\eta_{1}\ \\
\times\ g_{a}^{\ \,c}(\eta,\eta_{1})\ \gamma_{c}^{\ \,d_{1}d_{2}}(\vk, \vk_{1},\vk_{2})\ 
g_{d_{1}}^{\ \,b_{1}}(\eta_{1},\eta_{0})\ 
g_{d_{2}}^{\ \,b_{2}}(\eta_{1},\eta_{0}) \;,
\end{split}
\end{equation}
and
\begin{equation}
\begin{split}
{\Gamma}_{a}^{\ b_{1}b_{2}b_{3}}(\vk,\vk_{1},\vk_{2},\vk_{3};\eta,\eta_{0})=2
\int_{\eta_{0}}^{\eta}\dd\eta_{1}\int_{\eta_{0}}^{\eta_{1}}\dd\eta_{2}\ \\
\times\, g_{a}^{\ \,c}(\eta,\eta_{1})\ \gamma_{c}^{\ \,d_{1}d_{2}}(\vk_{1},\vk_{2}+\vk_{3})\ 
g_{d_{1}}^{\ \,b_{1}}(\eta_{1},\eta_{0}) \;
 g_{d_{2}}^{\ \,e}(\eta_{1},\eta_{2})\\
\times \, \gamma_{e}^{\ \,f_{2}f_{3}}(\vk_2 + \vk_3,\vk_{2},\vk_{3})\ 
g_{f_{2}}^{\ \,b_{2}}(\eta_{2},\eta_{0})\ 
g_{f_{3}}^{\ \,b_{3}}(\eta_{2},\eta_{0}) \;,
\end{split}
\end{equation}
which correspond to the diagram values of Fig.~\ref{MultiExpansPsi}.

\bibliography{PowerEikonal,IsodensityDamping_6Notes}

\begin{thebibliography}{10}%
\makeatletter
\providecommand \@ifxundefined [1]{%
 \ifx #1\undefined \expandafter \@firstoftwo
 \else \expandafter \@secondoftwo
\fi
}%
\providecommand \@ifnum [1]{%
 \ifnum #1\expandafter \@firstoftwo
 \else \expandafter \@secondoftwo
\fi
}%
\providecommand \enquote [1]{``#1''}%
\providecommand \bibnamefont  [1]{#1}%
\providecommand \bibfnamefont [1]{#1}%
\providecommand \citenamefont [1]{#1}%
\providecommand\href[0]{\@sanitize\@href}%
\providecommand\@href[1]{\endgroup\@@startlink{#1}\endgroup\@@href}%
\providecommand\@@href[1]{#1\@@endlink}%
\providecommand \@sanitize [0]{\begingroup\catcode`\&12\catcode`\#12\relax}%
\@ifxundefined \pdfoutput {\@firstoftwo}{%
 \@ifnum{\z@=\pdfoutput}{\@firstoftwo}{\@secondoftwo}%
}{%
 \providecommand\@@startlink[1]{\leavevmode\special{html:<a href="#1">}}%
 \providecommand\@@endlink[0]{\special{html:</a>}}%
}{%
 \providecommand\@@startlink[1]{%
  \leavevmode
  \pdfstartlink
   attr{/Border[0 0 1 ]/H/I/C[0 1 1]}%
   user{/Subtype/Link/A<</Type/Action/S/URI/URI(#1)>>}%
  \relax
 }%
 \providecommand\@@endlink[0]{\pdfendlink}%
}%
\providecommand \url  [0]{\begingroup\@sanitize \@url }%
\providecommand \@url [1]{\endgroup\@href {#1}{\urlprefix}}%
\providecommand \urlprefix [0]{URL }%
\providecommand \Eprint[0]{\href }%
\@ifxundefined \urlstyle {%
  \providecommand \doi [1]{doi:\discretionary{}{}{}#1}%
}{%
  \providecommand \doi [0]{doi:\discretionary{}{}{}\begingroup
  \urlstyle{rm}\Url }%
}%
\providecommand \doibase [0]{http://dx.doi.org/}%
\providecommand \Doi[1]{\href{\doibase#1}}%
\providecommand \bibAnnote [3]{%
  \BibitemShut{#1}%
  \begin{quotation}\noindent
    \textsc{Key:}\ #2\\\textsc{Annotation:}\ #3%
  \end{quotation}%
}%
\providecommand \bibAnnoteFile [2]{%
  \IfFileExists{#2}{\bibAnnote {#1} {#2} {\input{#2}}}{}%
}%
\providecommand \typeout [0]{\immediate \write \m@ne }%
\providecommand \selectlanguage [0]{\@gobble}%
\providecommand \bibinfo [0]{\@secondoftwo}%
\providecommand \bibfield [0]{\@secondoftwo}%
\providecommand \translation [1]{[#1]}%
\providecommand \BibitemOpen[0]{}%
\providecommand \bibitemStop [0]{}%
\providecommand \bibitemNoStop [0]{.\EOS\space}%
\providecommand \EOS [0]{\spacefactor3000\relax}%
\providecommand \BibitemShut [1]{\csname bibitem#1\endcsname}%
\bibitem{Crocce:2005xy}%
  \BibitemOpen
  \bibfield{author}{%
  \bibinfo {author} {\bibfnamefont{M.}~\bibnamefont{Crocce}}\ and\ \bibinfo
  {author} {\bibfnamefont{R.}~\bibnamefont{Scoccimarro}},\ }%
  \bibfield{journal}{%
  \Doi{10.1103/PhysRevD.73.063519}{\bibinfo {journal} {Phys. Rev.}}\ }%
  \textbf{\bibinfo {volume} {D73}},\ \bibinfo {pages} {063519} (\bibinfo {year}
  {2006}),\
  \Eprint{http://arxiv.org/abs/astro-ph/0509418}{arXiv:astro-ph/0509418}%
  \bibAnnoteFile{NoStop}{Crocce:2005xy}%
\bibitem{Crocce:2005xz}%
  \BibitemOpen
  \bibfield{author}{%
  \bibinfo {author} {\bibfnamefont{M.}~\bibnamefont{Crocce}}\ and\ \bibinfo
  {author} {\bibfnamefont{R.}~\bibnamefont{Scoccimarro}},\ }%
  \bibfield{journal}{%
  \Doi{10.1103/PhysRevD.73.063520}{\bibinfo {journal} {Phys. Rev.}}\ }%
  \textbf{\bibinfo {volume} {D73}},\ \bibinfo {pages} {063520} (\bibinfo {year}
  {2006}),\
  \Eprint{http://arxiv.org/abs/astro-ph/0509419}{arXiv:astro-ph/0509419}%
  \bibAnnoteFile{NoStop}{Crocce:2005xz}%
\bibitem{Crocce:2007dt}%
  \BibitemOpen
  \bibfield{author}{%
  \bibinfo {author} {\bibfnamefont{M.}~\bibnamefont{Crocce}}\ and\ \bibinfo
  {author} {\bibfnamefont{R.}~\bibnamefont{Scoccimarro}},\ }%
  \bibfield{journal}{%
  \Doi{10.1103/PhysRevD.77.023533}{\bibinfo {journal} {Phys. Rev.}}\ }%
  \textbf{\bibinfo {volume} {D77}},\ \bibinfo {pages} {023533} (\bibinfo {year}
  {2008}),\ \Eprint{http://arxiv.org/abs/0704.2783}{arXiv:0704.2783
  [astro-ph]}%
  \bibAnnoteFile{NoStop}{Crocce:2007dt}%
\bibitem{Matsubara:2007wj}%
  \BibitemOpen
  \bibfield{author}{%
  \bibinfo {author} {\bibfnamefont{T.}~\bibnamefont{Matsubara}},\ }%
  \bibfield{journal}{%
  \Doi{10.1103/PhysRevD.77.063530}{\bibinfo {journal} {Phys. Rev.}}\ }%
  \textbf{\bibinfo {volume} {D77}},\ \bibinfo {pages} {063530} (\bibinfo {year}
  {2008}),\ \Eprint{http://arxiv.org/abs/0711.2521}{arXiv:0711.2521
  [astro-ph]}%
  \bibAnnoteFile{NoStop}{Matsubara:2007wj}%
\bibitem{Matsubara:2008wx}%
  \BibitemOpen
  \bibfield{author}{%
  \bibinfo {author} {\bibfnamefont{T.}~\bibnamefont{Matsubara}},\ }%
  \bibfield{journal}{%
  \Doi{10.1103/PhysRevD.78.083519}{\bibinfo {journal} {Phys. Rev.}}\ }%
  \textbf{\bibinfo {volume} {D78}},\ \bibinfo {pages} {083519} (\bibinfo {year}
  {2008}),\ \Eprint{http://arxiv.org/abs/0807.1733}{arXiv:0807.1733
  [astro-ph]}%
  \bibAnnoteFile{NoStop}{Matsubara:2008wx}%
\bibitem{McDonald:2006hf}%
  \BibitemOpen
  \bibfield{author}{%
  \bibinfo {author} {\bibfnamefont{P.}~\bibnamefont{McDonald}},\ }%
  \bibfield{journal}{%
  \Doi{10.1103/PhysRevD.75.043514}{\bibinfo {journal} {Phys. Rev.}}\ }%
  \textbf{\bibinfo {volume} {D75}},\ \bibinfo {pages} {043514} (\bibinfo {year}
  {2007}),\
  \Eprint{http://arxiv.org/abs/astro-ph/0606028}{arXiv:astro-ph/0606028}%
  \bibAnnoteFile{NoStop}{McDonald:2006hf}%
\bibitem{Izumi:2007su}%
  \BibitemOpen
  \bibfield{author}{%
  \bibinfo {author} {\bibfnamefont{K.}~\bibnamefont{Izumi}}\ and\ \bibinfo
  {author} {\bibfnamefont{J.}~\bibnamefont{Soda}},\ }%
  \bibfield{journal}{%
  \Doi{10.1103/PhysRevD.76.083517}{\bibinfo {journal} {Phys. Rev.}}\ }%
  \textbf{\bibinfo {volume} {D76}},\ \bibinfo {pages} {083517} (\bibinfo {year}
  {2007}),\ \Eprint{http://arxiv.org/abs/0706.1604}{arXiv:0706.1604
  [astro-ph]}%
  \bibAnnoteFile{NoStop}{Izumi:2007su}%
\bibitem{Taruya:2007xy}%
  \BibitemOpen
  \bibfield{author}{%
  \bibinfo {author} {\bibfnamefont{A.}~\bibnamefont{Taruya}}\ and\ \bibinfo
  {author} {\bibfnamefont{T.}~\bibnamefont{Hiramatsu}},\ }%
  \bibfield{journal}{%
  \bibinfo {journal} {Astrophys.J.}\ }%
  \textbf{\bibinfo {volume} {674}},\ \bibinfo {pages} {617} (\bibinfo {year}
  {2008}),\ \Eprint{http://arxiv.org/abs/0708.1367}{arXiv:0708.1367
  [astro-ph]}%
  \bibAnnoteFile{NoStop}{Taruya:2007xy}%
\bibitem{Taruya:2009ir}%
  \BibitemOpen
  \bibfield{author}{%
  \bibinfo {author} {\bibfnamefont{A.}~\bibnamefont{Taruya}}, \bibinfo {author}
  {\bibfnamefont{T.}~\bibnamefont{Nishimichi}}, \bibinfo {author}
  {\bibfnamefont{S.}~\bibnamefont{Saito}},\ and\ \bibinfo {author}
  {\bibfnamefont{T.}~\bibnamefont{Hiramatsu}},\ }%
  \bibfield{journal}{%
  \Doi{10.1103/PhysRevD.80.123503}{\bibinfo {journal} {Phys. Rev.}}\ }%
  \textbf{\bibinfo {volume} {D80}},\ \bibinfo {pages} {123503} (\bibinfo {year}
  {2009}),\ \Eprint{http://arxiv.org/abs/0906.0507}{arXiv:0906.0507
  [astro-ph.CO]}%
  \bibAnnoteFile{NoStop}{Taruya:2009ir}%
\bibitem{Pietroni:2008jx}%
  \BibitemOpen
  \bibfield{author}{%
  \bibinfo {author} {\bibfnamefont{M.}~\bibnamefont{Pietroni}},\ }%
  \bibfield{journal}{%
  \Doi{10.1088/1475-7516/2008/10/036}{\bibinfo {journal} {JCAP}}\ }%
  \textbf{\bibinfo {volume} {0810}},\ \bibinfo {pages} {036} (\bibinfo {year}
  {2008}),\ \Eprint{http://arxiv.org/abs/0806.0971}{arXiv:0806.0971
  [astro-ph]}%
  \bibAnnoteFile{NoStop}{Pietroni:2008jx}%
\bibitem{Matarrese:2007wc}%
  \BibitemOpen
  \bibfield{author}{%
  \bibinfo {author} {\bibfnamefont{S.}~\bibnamefont{Matarrese}}\ and\ \bibinfo
  {author} {\bibfnamefont{M.}~\bibnamefont{Pietroni}},\ }%
  \bibfield{journal}{%
  \bibinfo {journal} {JCAP}\ }%
  \textbf{\bibinfo {volume} {0706}},\ \bibinfo {pages} {026} (\bibinfo {year}
  {2007}),\
  \Eprint{http://arxiv.org/abs/astro-ph/0703563}{arXiv:astro-ph/0703563}%
  \bibAnnoteFile{NoStop}{Matarrese:2007wc}%
\bibitem{Valageas:2003gm}%
  \BibitemOpen
  \bibfield{author}{%
  \bibinfo {author} {\bibfnamefont{P.}~\bibnamefont{Valageas}},\ }%
  \bibfield{journal}{%
  \Doi{10.1051/0004-6361:20040125}{\bibinfo {journal} {Astron. Astrophys.}}\ }%
  \textbf{\bibinfo {volume} {421}},\ \bibinfo {pages} {23} (\bibinfo {year}
  {2004}),\
  \Eprint{http://arxiv.org/abs/astro-ph/0307008}{arXiv:astro-ph/0307008}%
  \bibAnnoteFile{NoStop}{Valageas:2003gm}%
\bibitem{Valageas:2006bi}%
  \BibitemOpen
  \bibfield{author}{%
  \bibinfo {author} {\bibfnamefont{P.}~\bibnamefont{Valageas}},\ }%
  \bibfield{journal}{%
  \bibinfo {journal} {Astron. Astrophys.}\ }%
  \textbf{\bibinfo {volume} {465}},\ \bibinfo {pages} {725} (\bibinfo {year}
  {2007}),\
  \Eprint{http://arxiv.org/abs/astro-ph/0611849}{arXiv:astro-ph/0611849}%
  \bibAnnoteFile{NoStop}{Valageas:2006bi}%
\bibitem{2012arXiv1207.1465C}%
  \BibitemOpen
  \bibfield{author}{%
  \bibinfo {author} {\bibfnamefont{M.}~\bibnamefont{{Crocce}}}, \bibinfo
  {author} {\bibfnamefont{R.}~\bibnamefont{{Scoccimarro}}},\ and\ \bibinfo
  {author} {\bibfnamefont{F.}~\bibnamefont{{Bernardeau}}},\ }%
  \bibfield{journal}{%
  \bibinfo {journal} {ArXiv e-prints}}%
   (\bibinfo {month} {Jul.}\ \bibinfo {year} {2012}),\
  \Eprint{http://arxiv.org/abs/1207.1465}{arXiv:1207.1465 [astro-ph.CO]}%
  \bibAnnoteFile{NoStop}{2012arXiv1207.1465C}%
\bibitem{2012arXiv1208.1191T}%
  \BibitemOpen
  \bibfield{author}{%
  \bibinfo {author} {\bibfnamefont{A.}~\bibnamefont{{Taruya}}}, \bibinfo
  {author} {\bibfnamefont{F.}~\bibnamefont{{Bernardeau}}}, \bibinfo {author}
  {\bibfnamefont{T.}~\bibnamefont{{Nishimichi}}},\ and\ \bibinfo {author}
  {\bibfnamefont{S.}~\bibnamefont{{Codis}}},\ }%
  \bibfield{journal}{%
  \bibinfo {journal} {ArXiv e-prints}}%
   (\bibinfo {month} {Aug.}\ \bibinfo {year} {2012}),\
  \Eprint{http://arxiv.org/abs/1208.1191}{arXiv:1208.1191 [astro-ph.CO]}%
  \bibAnnoteFile{NoStop}{2012arXiv1208.1191T}%
\bibitem{2012JCAP...04..013T}%
  \BibitemOpen
  \bibfield{author}{%
  \bibinfo {author} {\bibfnamefont{S.}~\bibnamefont{{Tassev}}}\ and\ \bibinfo
  {author} {\bibfnamefont{M.}~\bibnamefont{{Zaldarriaga}}},\ }%
  \bibfield{journal}{%
  \Doi{10.1088/1475-7516/2012/04/013}{\bibinfo {journal} {\jcap}}\ }%
  \textbf{\bibinfo {volume} {4}},\ \bibinfo {eid} {013} (\bibinfo {month}
  {Apr.}\ \bibinfo {year} {2012}),\
  \Eprint{http://arxiv.org/abs/1109.4939}{arXiv:1109.4939 [astro-ph.CO]}%
  \bibAnnoteFile{NoStop}{2012JCAP...04..013T}%
\bibitem{2012arXiv1206.2926C}%
  \BibitemOpen
  \bibfield{author}{%
  \bibinfo {author} {\bibfnamefont{J.~J.~M.}\ \bibnamefont{{Carrasco}}},
  \bibinfo {author} {\bibfnamefont{M.~P.}\ \bibnamefont{{Hertzberg}}},\ and\
  \bibinfo {author} {\bibfnamefont{L.}~\bibnamefont{{Senatore}}},\ }%
  \bibfield{journal}{%
  \bibinfo {journal} {ArXiv e-prints}}%
   (\bibinfo {month} {Jun.}\ \bibinfo {year} {2012}),\
  \Eprint{http://arxiv.org/abs/1206.2926}{arXiv:1206.2926 [astro-ph.CO]}%
  \bibAnnoteFile{NoStop}{2012arXiv1206.2926C}%
\bibitem{Carlson:2009it}%
  \BibitemOpen
  \bibfield{author}{%
  \bibinfo {author} {\bibfnamefont{J.}~\bibnamefont{Carlson}}, \bibinfo
  {author} {\bibfnamefont{M.}~\bibnamefont{White}},\ and\ \bibinfo {author}
  {\bibfnamefont{N.}~\bibnamefont{Padmanabhan}},\ }%
  \bibfield{journal}{%
  \bibinfo {journal} {Phys. Rev.}\ }%
  \textbf{\bibinfo {volume} {D80}},\ \bibinfo {pages} {043531} (\bibinfo {year}
  {2009}),\ \Eprint{http://arxiv.org/abs/0905.0479}{arXiv:0905.0479
  [astro-ph.CO]}%
  \bibAnnoteFile{NoStop}{Carlson:2009it}%
\bibitem{2008PhRvD..78j3521B}%
  \BibitemOpen
  \bibfield{author}{%
  \bibinfo {author} {\bibfnamefont{F.}~\bibnamefont{{Bernardeau}}}, \bibinfo
  {author} {\bibfnamefont{M.}~\bibnamefont{{Crocce}}},\ and\ \bibinfo {author}
  {\bibfnamefont{R.}~\bibnamefont{{Scoccimarro}}},\ }%
  \bibfield{journal}{%
  \Doi{10.1103/PhysRevD.78.103521}{\bibinfo {journal} {\prd}}\ }%
  \textbf{\bibinfo {volume} {78}},\ \bibinfo {pages} {103521} (\bibinfo {month}
  {Nov.}\ \bibinfo {year} {2008}),\
  \Eprint{http://arxiv.org/abs/0806.2334}{arXiv:0806.2334}%
  \bibAnnoteFile{NoStop}{2008PhRvD..78j3521B}%
\bibitem{2010PhRvD..82h3507B}%
  \BibitemOpen
  \bibfield{author}{%
  \bibinfo {author} {\bibfnamefont{F.}~\bibnamefont{{Bernardeau}}}, \bibinfo
  {author} {\bibfnamefont{M.}~\bibnamefont{{Crocce}}},\ and\ \bibinfo {author}
  {\bibfnamefont{E.}~\bibnamefont{{Sefusatti}}},\ }%
  \bibfield{journal}{%
  \Doi{10.1103/PhysRevD.82.083507}{\bibinfo {journal} {\prd}}\ }%
  \textbf{\bibinfo {volume} {82}},\ \bibinfo {pages} {083507} (\bibinfo {month}
  {Oct.}\ \bibinfo {year} {2010}),\
  \Eprint{http://arxiv.org/abs/1006.4656}{arXiv:1006.4656 [astro-ph.CO]}%
  \bibAnnoteFile{NoStop}{2010PhRvD..82h3507B}%
\bibitem{2012PhRvD..85f3509B}%
  \BibitemOpen
  \bibfield{author}{%
  \bibinfo {author} {\bibfnamefont{F.}~\bibnamefont{{Bernardeau}}}, \bibinfo
  {author} {\bibfnamefont{N.}~\bibnamefont{{van de Rijt}}},\ and\ \bibinfo
  {author} {\bibfnamefont{F.}~\bibnamefont{{Vernizzi}}},\ }%
  \bibfield{journal}{%
  \Doi{10.1103/PhysRevD.85.063509}{\bibinfo {journal} {\prd}}\ }%
  \textbf{\bibinfo {volume} {85}},\ \bibinfo {eid} {063509} (\bibinfo {month}
  {Mar.}\ \bibinfo {year} {2012}),\
  \Eprint{http://arxiv.org/abs/1109.3400}{arXiv:1109.3400 [astro-ph.CO]}%
  \bibAnnoteFile{NoStop}{2012PhRvD..85f3509B}%
\bibitem{2008PhRvD..78h3503B}%
  \BibitemOpen
  \bibfield{author}{%
  \bibinfo {author} {\bibfnamefont{F.}~\bibnamefont{{Bernardeau}}}\ and\
  \bibinfo {author} {\bibfnamefont{P.}~\bibnamefont{{Valageas}}},\ }%
  \bibfield{journal}{%
  \Doi{10.1103/PhysRevD.78.083503}{\bibinfo {journal} {\prd}}\ }%
  \textbf{\bibinfo {volume} {78}},\ \bibinfo {pages} {083503} (\bibinfo {month}
  {Oct.}\ \bibinfo {year} {2008}),\
  \Eprint{http://arxiv.org/abs/0805.0805}{arXiv:0805.0805}%
  \bibAnnoteFile{NoStop}{2008PhRvD..78h3503B}%
\bibitem{2007arXiv0711.3407V}%
  \BibitemOpen
  \bibfield{author}{%
  \bibinfo {author} {\bibfnamefont{P.}~\bibnamefont{{Valageas}}},\ }%
  \bibfield{journal}{%
  \bibinfo {journal} {ArXiv e-prints}\ }%
  \textbf{\bibinfo {volume} {711}} (\bibinfo {month} {Nov.}\ \bibinfo {year}
  {2007}),\ \Eprint{http://arxiv.org/abs/0711.3407}{0711.3407}%
  \bibAnnoteFile{NoStop}{2007arXiv0711.3407V}%
\bibitem{2002PhR...372....1C}%
  \BibitemOpen
  \bibfield{author}{%
  \bibinfo {author} {\bibfnamefont{A.}~\bibnamefont{{Cooray}}}\ and\ \bibinfo
  {author} {\bibfnamefont{R.}~\bibnamefont{{Sheth}}},\ }%
  \bibfield{journal}{%
  \bibinfo {journal} {\physrep}\ }%
  \textbf{\bibinfo {volume} {372}},\ \bibinfo {pages} {1} (\bibinfo {month}
  {Dec.}\ \bibinfo {year} {2002}),\
  \Eprint{http://arxiv.org/abs/astro-ph/0206508}{astro-ph/0206508}%
  \bibAnnoteFile{NoStop}{2002PhR...372....1C}%
\bibitem{2010PhRvD..82h3520T}%
  \BibitemOpen
  \bibfield{author}{%
  \bibinfo {author} {\bibfnamefont{D.}~\bibnamefont{{Tseliakhovich}}}\ and\
  \bibinfo {author} {\bibfnamefont{C.}~\bibnamefont{{Hirata}}},\ }%
  \bibfield{journal}{%
  \Doi{10.1103/PhysRevD.82.083520}{\bibinfo {journal} {\prd}}\ }%
  \textbf{\bibinfo {volume} {82}},\ \bibinfo {pages} {083520} (\bibinfo {month}
  {Oct.}\ \bibinfo {year} {2010}),\
  \Eprint{http://arxiv.org/abs/1005.2416}{arXiv:1005.2416 [astro-ph.CO]}%
  \bibAnnoteFile{NoStop}{2010PhRvD..82h3520T}%
\bibitem{2011MNRAS.418..906T}%
  \BibitemOpen
  \bibfield{author}{%
  \bibinfo {author} {\bibfnamefont{D.}~\bibnamefont{{Tseliakhovich}}}, \bibinfo
  {author} {\bibfnamefont{R.}~\bibnamefont{{Barkana}}},\ and\ \bibinfo {author}
  {\bibfnamefont{C.~M.}\ \bibnamefont{{Hirata}}},\ }%
  \bibfield{journal}{%
  \Doi{10.1111/j.1365-2966.2011.19541.x}{\bibinfo {journal} {\mnras}}\ }%
  \textbf{\bibinfo {volume} {418}},\ \bibinfo {pages} {906} (\bibinfo {month}
  {Dec.}\ \bibinfo {year} {2011}),\
  \Eprint{http://arxiv.org/abs/1012.2574}{arXiv:1012.2574 [astro-ph.CO]}%
  \bibAnnoteFile{NoStop}{2011MNRAS.418..906T}%
\bibitem{Bovy:2012af}%
  \BibitemOpen
  \bibfield{author}{%
  \bibinfo {author} {\bibfnamefont{J.}~\bibnamefont{Bovy}}\ and\ \bibinfo
  {author} {\bibfnamefont{C.}~\bibnamefont{Dvorkin}}}%
   (\bibinfo {year} {2012}),\
  \Eprint{http://arxiv.org/abs/1205.2083}{arXiv:1205.2083 [astro-ph.CO]}%
  \bibAnnoteFile{NoStop}{Bovy:2012af}%
\bibitem{2011MNRAS.412L..40M}%
  \BibitemOpen
  \bibfield{author}{%
  \bibinfo {author} {\bibfnamefont{U.}~\bibnamefont{{Maio}}}, \bibinfo {author}
  {\bibfnamefont{L.~V.~E.}\ \bibnamefont{{Koopmans}}},\ and\ \bibinfo {author}
  {\bibfnamefont{B.}~\bibnamefont{{Ciardi}}},\ }%
  \bibfield{journal}{%
  \Doi{10.1111/j.1745-3933.2010.01001.x}{\bibinfo {journal} {\mnras}}\ }%
  \textbf{\bibinfo {volume} {412}},\ \bibinfo {pages} {L40} (\bibinfo {month}
  {Mar.}\ \bibinfo {year} {2011}),\
  \Eprint{http://arxiv.org/abs/1011.4006}{arXiv:1011.4006 [astro-ph.CO]}%
  \bibAnnoteFile{NoStop}{2011MNRAS.412L..40M}%
\bibitem{2011ApJ...730L...1S}%
  \BibitemOpen
  \bibfield{author}{%
  \bibinfo {author} {\bibfnamefont{A.}~\bibnamefont{{Stacy}}}, \bibinfo
  {author} {\bibfnamefont{V.}~\bibnamefont{{Bromm}}},\ and\ \bibinfo {author}
  {\bibfnamefont{A.}~\bibnamefont{{Loeb}}},\ }%
  \bibfield{journal}{%
  \Doi{10.1088/2041-8205/730/1/L1}{\bibinfo {journal} {\apjl}}\ }%
  \textbf{\bibinfo {volume} {730}},\ \bibinfo {eid} {L1} (\bibinfo {month}
  {Mar.}\ \bibinfo {year} {2011}),\
  \Eprint{http://arxiv.org/abs/1011.4512}{arXiv:1011.4512 [astro-ph.CO]}%
  \bibAnnoteFile{NoStop}{2011ApJ...730L...1S}%
\bibitem{2011ApJ...736..147G}%
  \BibitemOpen
  \bibfield{author}{%
  \bibinfo {author} {\bibfnamefont{T.~H.}\ \bibnamefont{{Greif}}}, \bibinfo
  {author} {\bibfnamefont{S.~D.~M.}\ \bibnamefont{{White}}}, \bibinfo {author}
  {\bibfnamefont{R.~S.}\ \bibnamefont{{Klessen}}},\ and\ \bibinfo {author}
  {\bibfnamefont{V.}~\bibnamefont{{Springel}}},\ }%
  \bibfield{journal}{%
  \Doi{10.1088/0004-637X/736/2/147}{\bibinfo {journal} {\apj}}\ }%
  \textbf{\bibinfo {volume} {736}},\ \bibinfo {eid} {147} (\bibinfo {month}
  {Aug.}\ \bibinfo {year} {2011}),\
  \Eprint{http://arxiv.org/abs/1101.5493}{arXiv:1101.5493 [astro-ph.CO]}%
  \bibAnnoteFile{NoStop}{2011ApJ...736..147G}%
\bibitem{2011arXiv1110.2111F}%
  \BibitemOpen
  \bibfield{author}{%
  \bibinfo {author} {\bibfnamefont{A.}~\bibnamefont{{Fialkov}}}, \bibinfo
  {author} {\bibfnamefont{R.}~\bibnamefont{{Barkana}}}, \bibinfo {author}
  {\bibfnamefont{D.}~\bibnamefont{{Tseliakhovich}}},\ and\ \bibinfo {author}
  {\bibfnamefont{C.~M.}\ \bibnamefont{{Hirata}}},\ }%
  \bibfield{journal}{%
  \bibinfo {journal} {ArXiv e-prints}}%
   (\bibinfo {month} {Oct.}\ \bibinfo {year} {2011}),\
  \Eprint{http://arxiv.org/abs/1110.2111}{arXiv:1110.2111 [astro-ph.CO]}%
  \bibAnnoteFile{NoStop}{2011arXiv1110.2111F}%
\bibitem{Bittner:2011rx}%
  \BibitemOpen
  \bibfield{author}{%
  \bibinfo {author} {\bibfnamefont{J.~M.}\ \bibnamefont{Bittner}}\ and\
  \bibinfo {author} {\bibfnamefont{A.}~\bibnamefont{Loeb}}}%
   (\bibinfo {year} {2011}),\
  \Eprint{http://arxiv.org/abs/1110.4659}{arXiv:1110.4659 [astro-ph.CO]}%
  \bibAnnoteFile{NoStop}{Bittner:2011rx}%
\bibitem{McQuinn:2012rt}%
  \BibitemOpen
  \bibfield{author}{%
  \bibinfo {author} {\bibfnamefont{M.}~\bibnamefont{McQuinn}}\ and\ \bibinfo
  {author} {\bibfnamefont{R.~M.}\ \bibnamefont{O'Leary}}}%
   (\bibinfo {year} {2012}),\
  \Eprint{http://arxiv.org/abs/1204.1345}{arXiv:1204.1345 [astro-ph.CO]}%
  \bibAnnoteFile{NoStop}{McQuinn:2012rt}%
\bibitem{2010JCAP...11..007D}%
  \BibitemOpen
  \bibfield{author}{%
  \bibinfo {author} {\bibfnamefont{N.}~\bibnamefont{{Dalal}}}, \bibinfo
  {author} {\bibfnamefont{U.-L.}\ \bibnamefont{{Pen}}},\ and\ \bibinfo {author}
  {\bibfnamefont{U.}~\bibnamefont{{Seljak}}},\ }%
  \bibfield{journal}{%
  \Doi{10.1088/1475-7516/2010/11/007}{\bibinfo {journal} {\jcap}}\ }%
  \textbf{\bibinfo {volume} {11}},\ \bibinfo {pages} {7} (\bibinfo {month}
  {Nov.}\ \bibinfo {year} {2010}),\
  \Eprint{http://arxiv.org/abs/1009.4704}{arXiv:1009.4704 [astro-ph.CO]}%
  \bibAnnoteFile{NoStop}{2010JCAP...11..007D}%
\bibitem{2011JCAP...07..018Y}%
  \BibitemOpen
  \bibfield{author}{%
  \bibinfo {author} {\bibfnamefont{J.}~\bibnamefont{{Yoo}}}, \bibinfo {author}
  {\bibfnamefont{N.}~\bibnamefont{{Dalal}}},\ and\ \bibinfo {author}
  {\bibfnamefont{U.}~\bibnamefont{{Seljak}}},\ }%
  \bibfield{journal}{%
  \Doi{10.1088/1475-7516/2011/07/018}{\bibinfo {journal} {\jcap}}\ }%
  \textbf{\bibinfo {volume} {7}},\ \bibinfo {pages} {18} (\bibinfo {month}
  {Jul.}\ \bibinfo {year} {2011}),\
  \Eprint{http://arxiv.org/abs/1105.3732}{arXiv:1105.3732 [astro-ph.CO]}%
  \bibAnnoteFile{NoStop}{2011JCAP...07..018Y}%
\bibitem{Ferraro:2011nc}%
  \BibitemOpen
  \bibfield{author}{%
  \bibinfo {author} {\bibfnamefont{S.}~\bibnamefont{Ferraro}}, \bibinfo
  {author} {\bibfnamefont{K.~M.}\ \bibnamefont{Smith}},\ and\ \bibinfo {author}
  {\bibfnamefont{C.}~\bibnamefont{Dvorkin}},\ }%
  \bibfield{journal}{%
  \Doi{10.1103/PhysRevD.85.043523}{\bibinfo {journal} {Phys.Rev.}}\ }%
  \textbf{\bibinfo {volume} {D85}},\ \bibinfo {pages} {043523} (\bibinfo {year}
  {2012}),\ \Eprint{http://arxiv.org/abs/1110.2182}{arXiv:1110.2182
  [astro-ph.CO]}%
  \bibAnnoteFile{NoStop}{Ferraro:2011nc}%
\bibitem{Grin:2011nk}%
  \BibitemOpen
  \bibfield{author}{%
  \bibinfo {author} {\bibfnamefont{D.}~\bibnamefont{Grin}}, \bibinfo {author}
  {\bibfnamefont{O.}~\bibnamefont{Dore}},\ and\ \bibinfo {author}
  {\bibfnamefont{M.}~\bibnamefont{Kamionkowski}},\ }%
  \bibfield{journal}{%
  \Doi{10.1103/PhysRevLett.107.261301}{\bibinfo {journal} {Phys.Rev.Lett.}}\ }%
  \textbf{\bibinfo {volume} {107}},\ \bibinfo {pages} {261301} (\bibinfo {year}
  {2011}),\ \Eprint{http://arxiv.org/abs/1107.1716}{arXiv:1107.1716
  [astro-ph.CO]}%
  \bibAnnoteFile{NoStop}{Grin:2011nk}%
\bibitem{2002PhR...367....1B}%
  \BibitemOpen
  \bibfield{author}{%
  \bibinfo {author} {\bibfnamefont{F.}~\bibnamefont{{Bernardeau}}}, \bibinfo
  {author} {\bibfnamefont{S.}~\bibnamefont{{Colombi}}}, \bibinfo {author}
  {\bibfnamefont{E.}~\bibnamefont{{Gazta{\~n}aga}}},\ and\ \bibinfo {author}
  {\bibfnamefont{R.}~\bibnamefont{{Scoccimarro}}},\ }%
  \bibfield{journal}{%
  \bibinfo {journal} {\physrep}\ }%
  \textbf{\bibinfo {volume} {367}},\ \bibinfo {pages} {1} (\bibinfo {month}
  {Sep.}\ \bibinfo {year} {2002})%
  \bibAnnoteFile{NoStop}{2002PhR...367....1B}%
\bibitem{2010PhRvD..81b3524S}%
  \BibitemOpen
  \bibfield{author}{%
  \bibinfo {author} {\bibfnamefont{G.}~\bibnamefont{{Somogyi}}}\ and\ \bibinfo
  {author} {\bibfnamefont{R.~E.}\ \bibnamefont{{Smith}}},\ }%
  \bibfield{journal}{%
  \Doi{10.1103/PhysRevD.81.023524}{\bibinfo {journal} {\prd}}\ }%
  \textbf{\bibinfo {volume} {81}},\ \bibinfo {pages} {023524} (\bibinfo {month}
  {Jan.}\ \bibinfo {year} {2010}),\
  \Eprint{http://arxiv.org/abs/0910.5220}{arXiv:0910.5220 [astro-ph.CO]}%
  \bibAnnoteFile{NoStop}{2010PhRvD..81b3524S}%
\bibitem{2001NYASA.927...13S}%
  \BibitemOpen
  \bibfield{author}{%
  \bibinfo {author} {\bibfnamefont{R.}~\bibnamefont{{Scoccimarro}}},\ }%
  in\ \emph{\bibinfo {booktitle} {The Onset of Nonlinearity in Cosmology}},\
  \bibinfo {series} {New York Academy Sciences Annals}, Vol.\ \bibinfo {volume}
  {927},\ \bibinfo {editor} {edited by\ \bibinfo {editor}
  {\bibfnamefont{J.~N.}\ \bibnamefont{{Fry}}}, \bibinfo {editor}
  {\bibfnamefont{J.~R.}\ \bibnamefont{{Buchler}}},\ and\ \bibinfo {editor}
  {\bibfnamefont{H.}~\bibnamefont{{Kandrup}}}}\ (\bibinfo {year} {2001})\ pp.\
  \bibinfo {pages} {13--+}%
  \bibAnnoteFile{NoStop}{2001NYASA.927...13S}%
\bibitem{Lewis:1999bs}%
  \BibitemOpen
  \bibfield{author}{%
  \bibinfo {author} {\bibfnamefont{A.}~\bibnamefont{Lewis}}, \bibinfo {author}
  {\bibfnamefont{A.}~\bibnamefont{Challinor}},\ and\ \bibinfo {author}
  {\bibfnamefont{A.}~\bibnamefont{Lasenby}},\ }%
  \bibfield{journal}{%
  \bibinfo {journal} {Astrophys. J.}\ }%
  \textbf{\bibinfo {volume} {538}},\ \bibinfo {pages} {473} (\bibinfo {year}
  {2000}),\ \Eprint{http://arxiv.org/abs/astro-ph/9911177}{astro-ph/9911177}%
  \bibAnnoteFile{NoStop}{Lewis:1999bs}%
\bibitem{Chisari:2011iq}%
  \BibitemOpen
  \bibfield{author}{%
  \bibinfo {author} {\bibfnamefont{N.~E.}\ \bibnamefont{Chisari}}\ and\
  \bibinfo {author} {\bibfnamefont{M.}~\bibnamefont{Zaldarriaga}},\ }%
  \bibfield{journal}{%
  \Doi{10.1103/PhysRevD.84.089901, 10.1103/PhysRevD.83.123505}{\bibinfo
  {journal} {Phys.Rev.}}\ }%
  \textbf{\bibinfo {volume} {D83}},\ \bibinfo {pages} {123505} (\bibinfo {year}
  {2011}),\ \Eprint{http://arxiv.org/abs/1101.3555}{arXiv:1101.3555
  [astro-ph.CO]}%
  \bibAnnoteFile{NoStop}{Chisari:2011iq}%
\bibitem{Note1}%
  \BibitemOpen
  \bibinfo {note} {Note that the constant isodensity mode does not contribute
  to loops simply because $\gamma _{\alpha }^{\ {\triangleleft }\beta }
  ({\protect \bf k}, {\protect \bf q}, {\protect \bf k}) = 0$ at lowest order
  in $q/k$.}%
  \bibAnnoteFile{Stop}{Note1}%
\bibitem{Note2}%
  \BibitemOpen
  \bibinfo {note} {The average propagator ${\protect \mathaccentV {hat}05E{\xi
  }}_a^{\ b}(k, \eta ,\eta _0)$ coincides with what is usually defined as the
  nonlinear propagator $G_{ab}(k, \eta ,\eta _0)$ in \cite
  {2008PhRvD..78j3521B,2010PhRvD..82h3507B,2012PhRvD..85f3509B}.}%
  \bibAnnoteFile{Stop}{Note2}%
\bibitem{Note3}%
  \BibitemOpen
  \bibinfo {note} {Although we present the calculations for Gaussian initial
  conditions, the results derived here are valid irrespectively of the nature
  of the initial conditions, as our arguments do not depend on where the
  incoming line emerge from.}%
  \bibAnnoteFile{Stop}{Note3}%
\bibitem{Note4}%
  \BibitemOpen
  \bibinfo {note} {In \cite {2012arXiv1205.2235A}, a scaling relation has been
  put forward where the nonlinear power spectrum depends only on the reduced
  wave-mode $y \equiv e^{\eta } \sigma _d k$. While this definition
  approximately captures the time dependence of the power spectrum, its $\sigma
  _d$ dependence is in contradiction with our findings.}%
  \bibAnnoteFile{Stop}{Note4}%
\bibitem{Bernardeauetal1012a}%
  \BibitemOpen
  \bibfield{author}{%
  \bibinfo {author} {\bibfnamefont{F.}~\bibnamefont{{Bernardeau}}}, \bibinfo
  {author} {\bibfnamefont{A.}~\bibnamefont{{Taruya}}},\ and\ \bibinfo {author}
  {\bibfnamefont{T.}~\bibnamefont{{Nishimichi}}},\ }%
  \bibfield{journal}{%
  \bibinfo {journal} {in preparation}}%
   (\bibinfo {year} {2012})%
  \bibAnnoteFile{NoStop}{Bernardeauetal1012a}%
\bibitem{1980PhRvL..45.1980B}%
  \BibitemOpen
  \bibfield{author}{%
  \bibinfo {author} {\bibfnamefont{J.~R.}\ \bibnamefont{{Bond}}}, \bibinfo
  {author} {\bibfnamefont{G.}~\bibnamefont{{Efstathiou}}},\ and\ \bibinfo
  {author} {\bibfnamefont{J.}~\bibnamefont{{Silk}}},\ }%
  \bibfield{journal}{%
  \Doi{10.1103/PhysRevLett.45.1980.2}{\bibinfo {journal} {Physical Review
  Letters}}\ }%
  \textbf{\bibinfo {volume} {45}},\ \bibinfo {pages} {1980} (\bibinfo {month}
  {Dec.}\ \bibinfo {year} {1980})%
  \bibAnnoteFile{NoStop}{1980PhRvL..45.1980B}%
\bibitem{Note5}%
  \BibitemOpen
  \bibinfo {note} {It can be noted that the total density contrast in Fig.~\ref
  {EikonalDamping} can be occasionally below this asymptotic curve. This is due
  to the fact that the CDM and baryon density contrasts are transitorily in
  opposition of phase.}%
  \bibAnnoteFile{Stop}{Note5}%
\bibitem{1993ApJ...419....1K}%
  \BibitemOpen
  \bibfield{author}{%
  \bibinfo {author} {\bibfnamefont{A.}~\bibnamefont{{Kogut}}}, \bibinfo
  {author} {\bibfnamefont{C.}~\bibnamefont{{Lineweaver}}}, \bibinfo {author}
  {\bibfnamefont{G.~F.}\ \bibnamefont{{Smoot}}}, \bibinfo {author}
  {\bibfnamefont{C.~L.}\ \bibnamefont{{Bennett}}}, \bibinfo {author}
  {\bibfnamefont{A.}~\bibnamefont{{Banday}}}, \bibinfo {author}
  {\bibfnamefont{N.~W.}\ \bibnamefont{{Boggess}}}, \bibinfo {author}
  {\bibfnamefont{E.~S.}\ \bibnamefont{{Cheng}}}, \bibinfo {author}
  {\bibfnamefont{G.}~\bibnamefont{{de Amici}}}, \bibinfo {author}
  {\bibfnamefont{D.~J.}\ \bibnamefont{{Fixsen}}}, \bibinfo {author}
  {\bibfnamefont{G.}~\bibnamefont{{Hinshaw}}}, \bibinfo {author}
  {\bibfnamefont{P.~D.}\ \bibnamefont{{Jackson}}}, \bibinfo {author}
  {\bibfnamefont{M.}~\bibnamefont{{Janssen}}}, \bibinfo {author}
  {\bibfnamefont{P.}~\bibnamefont{{Keegstra}}}, \bibinfo {author}
  {\bibfnamefont{K.}~\bibnamefont{{Loewenstein}}}, \bibinfo {author}
  {\bibfnamefont{P.}~\bibnamefont{{Lubin}}}, \bibinfo {author}
  {\bibfnamefont{J.~C.}\ \bibnamefont{{Mather}}}, \bibinfo {author}
  {\bibfnamefont{L.}~\bibnamefont{{Tenorio}}}, \bibinfo {author}
  {\bibfnamefont{R.}~\bibnamefont{{Weiss}}}, \bibinfo {author}
  {\bibfnamefont{D.~T.}\ \bibnamefont{{Wilkinson}}},\ and\ \bibinfo {author}
  {\bibfnamefont{E.~L.}\ \bibnamefont{{Wright}}},\ }%
  \bibfield{journal}{%
  \Doi{10.1086/173453}{\bibinfo {journal} {\apj}}\ }%
  \textbf{\bibinfo {volume} {419}},\ \bibinfo {pages} {1} (\bibinfo {month}
  {Dec.}\ \bibinfo {year} {1993}),\
  \Eprint{http://arxiv.org/abs/arXiv:astro-ph/9312056}{arXiv:astro-ph/9312056}%
  \bibAnnoteFile{NoStop}{1993ApJ...419....1K}%
\bibitem{2012arXiv1205.2235A}%
  \BibitemOpen
  \bibfield{author}{%
  \bibinfo {author} {\bibfnamefont{S.}~\bibnamefont{{Anselmi}}}\ and\ \bibinfo
  {author} {\bibfnamefont{M.}~\bibnamefont{{Pietroni}}},\ }%
  \bibfield{journal}{%
  \bibinfo {journal} {ArXiv e-prints}}%
   (\bibinfo {month} {May}\ \bibinfo {year} {2012}),\
  \Eprint{http://arxiv.org/abs/1205.2235}{arXiv:1205.2235 [astro-ph.CO]}%
  \bibAnnoteFile{NoStop}{2012arXiv1205.2235A}%
\end{thebibliography}%

\end{document}